\documentclass{LMCS}

\def\dOi{9(4:19)2013}
\lmcsheading%
{\dOi}
{1--22}
{}
{}
{Feb.~12, 2013}
{Dec.~11, 2013}
{}

\ACMCCS{[{\bf Theory of computation}]: Models of
  computation-Computability---Recursive functions}

\usepackage{hyperref}
\usepackage{graphicx}

\theoremstyle{plain}

\newcommand{\diam}{\mathop{\mathrm{diam}}}

\newcommand{\mesh}{\mathop{\mathrm{mesh}}}
\newcommand{\fdiam}{\mathop{\mathrm{fdiam}}}
\newcommand{\fmesh}{\mathop{\mathrm{fmesh}}}
\newcommand{\Int}{\mathop{\mathrm{Int}}}
\newcommand{\Bd}{\mathop{\mathrm{Bd}}}

\begin{document}

\title[Compact manifolds with computable boundaries]{Compact manifolds
  with computable boundaries}

\author[Z.~Iljazovi\'{c}]{Zvonko Iljazovi\'{c}}   %required
\address{Department of Mathematics, Faculty of Science, University of Zagreb, Croatia} %required
\email{zilj@math.hr}  %optional
%\thanks{thanks 1, optional.}   %optional

%% etc.

%% required for running head on odd and even pages, use suitable
%% abbreviations in case of long titles and many authors:

%% mandatory lists of keywords and classifications:
%\subjclass{MANDATORY list of acm classifications}
%\titlecomment{OPTIONAL comment concerning the title, \eg, if a variant
%or an extended abstract of the paper has appeared elsewehere}
%%%%%%%%%%%%%%%%%%%%%%%%%%%%%%%%%%%%%%%%%%%%%%%%%%%%%%%%%%%%%%%%%%%%%%%%%%%

%% the abstract has to PRECEED the command \maketitle:
%% be sure not to issue the \maketitle command twice!

\begin{abstract}
  \noindent    We investigate conditions under which a co-computably enumerable closed set in
  a computable metric space is computable
  and prove that in each locally computable computable metric space each
  co-computably enumerable compact manifold with computable
  boundary
  is computable. In fact, we examine the notion of a semi-computable compact set and we prove a more general result: in any
  computable metric space each semi-computable compact manifold with computable boundary is
  computable. In particular, each semi-computable compact
  (boundaryless) manifold is computable.
\end{abstract}

\maketitle

%% start the paper here:
\section{Introduction}\label{introd}
If $f:\mathbb{R}\rightarrow \mathbb{R}$ is a computable function
and if for some $a<b$ we have $f(a)<0$ and $f(b)>0$, then there
exists a computable number $c$ between $a$ and $b$ such that
$f(c)=0$ \cite{mpir}. In general, however, we cannot conclude that
a computable function which has a zero-point has a computable
zero-point: there exists a computable function
$f:\mathbb{R}\rightarrow \mathbb{R}$ which has zero-points, but
none of them is computable \cite{specker}. However, it is  known
that if a computable function $f:\mathbb{R}^{m} \rightarrow
\mathbb{R}$, $m\geq 1$, has an isolated zero-point, then that
point must be computable. This raises the following question:
under what conditions a computable function $f:\mathbb{R}^{m}
\rightarrow \mathbb{R}$ has a computable zero-point, or, even
better, under what conditions the set $f^{-1}(\{0\})$ of all zero
points of $f$ is computable?

A closed subset of $\mathbb{R}^{m} $ is computable if it can be
effectively approximated by a finite set of points with rational
coordinates with arbitrary precision on an arbitrary  bounded
region of $\mathbb{R}^{m} $. Each nonempty computable set contains
computable points, in fact they are dense in it. On the other
hand, if $S\subseteq \mathbb{R}^{m} $, then $S$ is equal to
$f^{-1} (\{0\})$ for some computable function
$f:\mathbb{R}^{m}\rightarrow \mathbb{R}$ if and only if the
complement of $S$ can be effectively covered by open balls. A
closed subset of $\mathbb{R}^{m} $ is called co-computably
enumerable (co-c.e.) if its complement can be effectively covered
by open balls. So the question under what conditions a set of the
form $f^{-1} (\{0\})$ is computable can be restated in the
following way: under what conditions a co-c.e.\ set  is
computable, i.e.\ under what conditions the implication
\begin{equation}\label{intro-2}
S\mbox{ co-computably enumerable closed set } \Rightarrow S \mbox{
computable closed set}
\end{equation}
holds?

The implication (\ref{intro-2}) does not hold in general,
moreover, as  mentioned, there exists a nonempty co-c.e.\ subset
of $\mathbb{R}$ which contains no computable points. Such a set
cannot be connected (otherwise it would consist of a single point
or it would contain an open interval). It is easy to construct a
connected co-c.e.\ subset of $\mathbb{R}^{2}$ which contains no
computable points, but it is an interesting question is it
possible to achieve that such a set is even simply connected
\cite{ziegler}. Kihara  \cite{kihara} recently showed that such a
set does exist.

Regarding the implication (\ref{intro-2}), it fails to be true
even for sets which are very simple from the topological
viewpoint: in each $\mathbb{R}^{m} $ there exists a line segment
which is co-c.e., but not computable \cite{miller}. However, there
are certain conditions under which this implication holds. In
\cite{jucs} it was proved that (\ref{intro-2}) holds when $S$ is a
circularly chainable continuum or a continuum chainable from $a$
to $b$, where $a$ and $b$ are computable points. Brattka
\cite{brat} gave some results concerning (\ref{intro-2}) in the
case when $S$ is the graph of a continuous function. Miller
\cite{miller} proved that (\ref{intro-2}) holds
 whenever $S\subseteq \mathbb{R}^{m} $ is a topological sphere (i.e.\ homeomorphic to
 the unit sphere $S^{n}\subseteq \mathbb{R}^{n+1}$ for some $n$)
 or $S$ is homeomorphic to the closed unit ball $B^{n}\subseteq \mathbb{R}^{n}$ for some $n$
(i.e.\ $S$ is an $n-$cell) by a homeomorphism $f:B^{n} \rightarrow
S$ such that $f(S^{n-1})$
 is co-c.e.\ set in $\mathbb{R}^{m} $.
Furthermore, by \cite{lmcs}, these results hold not just in
$\mathbb{R}^{m}$, but  also in any
 computable metric space which is locally computable.

This, for example, means that each co-c.e.\ arc in $\mathbb{R}^{m}
$ is computable if its endpoints are computable. Hence the
computability of endpoints implies the computability of the whole
arc. And if we have a co-c.e.\ cell, then the computability of its
boundary sphere implies the computability of the whole cell. Since
arcs and cells are examples of spaces known as manifolds with
boundary, the question which arises here is what can be said about
computability of such spaces, i.e.\ if we have a co-c.e.\ manifold
with boundary, does the computability of its boundary imply the
computability of the whole manifold?

 In this paper we answer this question. We prove that if $(X,d,\alpha )$ is a locally
computable computable metric
 space and if
 $S$ is a co-c.e.\ closed set in $(X,d,\alpha )$
which is, as a subspace of $(X,d)$, a compact manifold with
boundary, then the computability of $\partial S$  implies the
computability of $S$.  This in particular means that a co-c.e.\
set $S$ is computable if it is a (boundaryless) manifold. A
manifold with boundary is a topological space in which each point
is contained in some open set which is homeomorphic to Euclidean
space $\mathbb{R}^{n} $ or Euclidean half-space $\{(x_{1} ,\dots
,x_{n} )\mid x_{n} \geq 0\}$. If $X$ is a manifold with boundary,
then the boundary $\partial X$ of $X$ consists of all points $x\in
X$ such that none of the open sets containing $x$ is homeomorphic
to Euclidean space. If $X$ is a manifold with boundary and
$\partial X=\emptyset $, than we simply say that $X$ is a
manifold.

For example, the unit sphere $S^{n}$ is a manifold and the unit
ball $B^{n} $ is a manifold with boundary, its boundary is
$S^{n-1}$. Moreover, each topological sphere is a manifold and if
$f:B^{n} \rightarrow X$ is a homeomorphism, then $X$ is a manifold
with boundary and $\partial X=f(S^{n-1})$.

In fact, we will prove a result which turns out to be more general:
in any computable metric space $(X,d,\alpha )$ the implication
\begin{equation}\label{intro-21}
S\mbox{ semi-computable compact set } \Rightarrow S \mbox{
computable compact set}
\end{equation}
holds if $S$ is a compact manifold with computable boundary. That
$S$ is a semi-computable compact set means that $S$ is compact and
we can effectively enumerate all rational open sets which cover
$S$. In other words, we will prove that if $S$ is semi-computable
compact manifold with boundary in $(X,d,\alpha )$ then the
following implication holds: $$\partial S\mbox{ computable }
\Rightarrow S \mbox{ computable}.$$

In order to prove this, the central notion will be the notion of
computability up to some set. Using techniques from \cite{jucs}
and \cite{lmcs} (in particular using $n-$dimensional chains) we
will see that in each semi-computable compact manifold with
computable boundary each point has a neighborhood which is
computable up to the manifold. This fact will then imply  that a
semi-computable compact manifold with computable boundary is
computable.

In Section \ref{prelim} we give some basic notions and in Section
\ref{sect-3} we examine semi-computable compact sets. In Section
\ref{sect-4} we introduce the notion of computability up to a set.
In Section \ref{sect-5} we examine $n$-chains and prove that in
each semi-computable compact manifold with computable boundary
each point has a neighborhood which is computable up to the
manifold (Theorem \ref{lok-izr-eukld-skupa}). In
 Section \ref{sect-6} we prove that semi-computable compact manifolds with computable boundaries are computable.

\section{Basic notions and techniques}\label{prelim}
If $X$ is a set, let $\mathcal{P}(X)$ denote the set of all
subsets of $X$.

For $m\in \mathbb{N}$ let  $\mathbb{N}_{m} =\{0,\dots ,m\}$. For
$n\geq 1$ let $$\mathbb{N}_{m} ^{n} =\{(x_{1} ,\dots ,x_{n} )\mid
x_{1} ,\dots ,x_{n}\in \mathbb{N}_{m} \}.$$

We say that a function $\Phi :\mathbb{N}^{k} \rightarrow
\mathcal{P}(\mathbb{N}^{n} )$ is \textbf{computable} if the
function $\overline{\Phi }:\mathbb{N}^{k+n}\rightarrow \mathbb{N}$
defined by
$$\overline{\Phi }(x,y)=\chi _{\Phi (x)}(y),$$ $x\in \mathbb{N}^{k} ,$ $y\in \mathbb{N}^{n}$,
is computable (i.e.\ recursive). Here  $\chi _{S}:\mathbb{N}^{n}
\rightarrow \{0,1\}$ denotes the characteristic function of
$S\subseteq \mathbb{N}^{n} $. A function $\Phi :\mathbb{N}^{k}
\rightarrow \mathcal{P}(\mathbb{N}^{n} )$ is said to be
\textbf{computably bounded} if there exists a computable function
$\varphi :\mathbb{N}^{k} \rightarrow \mathbb{N}$ such that $\Phi
(x)\subseteq \mathbb{N}_{\varphi (x)}^{n} $  for all $x\in
\mathbb{N}^{k}$.

We say that a function $\Phi :\mathbb{N}^{k} \rightarrow
\mathcal{P}(\mathbb{N}^{n} )$ is  \textbf{c.c.b}$.$  if $\Phi $ is
computable and computably bounded.

\begin{prop} \label{p1}\hfill
\begin{enumerate}%[\em(1)]
\item If $\Phi ,\Psi:\mathbb{N}^{k} \rightarrow
\mathcal{P}(\mathbb{N}^{n} )$ are c.c.b.\ functions, then the
function $\mathbb{N}^{k} \rightarrow \mathcal{P}(\mathbb{N}^{n}
)$, $x\mapsto \Phi (x)\cup \Psi (x)$ is c.c.b.

\item If $\Phi ,\Psi:\mathbb{N}^{k} \rightarrow
\mathcal{P}(\mathbb{N}^{n} )$ are c.c.b.\ functions, then the sets
$\{x\in \mathbb{N}^{k} \mid \Phi (x)=\Psi (x)\}$ and $\{x\in
\mathbb{N}^{k} \mid \Phi (x)\subseteq \Psi (x)\}$ are decidable.

\item Let $\Phi :\mathbb{N}^{k} \rightarrow
\mathcal{P}(\mathbb{N}^{n} )$ and $\Psi :\mathbb{N}^{n}\rightarrow
\mathcal{P}(\mathbb{N}^{m} )$ be c.c.b.\ functions. Let $\Lambda
:\mathbb{N}^{k} \rightarrow \mathcal{P}(\mathbb{N}^{m} )$ be
defined by $$\Lambda (x)=\bigcup_{z\in \Phi (x)}\Psi  (z),$$ $x\in
\mathbb{N}^{k} $. Then $\Lambda  $ is a c.c.b.\ function.

\item Let $\Phi :\mathbb{N}^{k} \rightarrow
\mathcal{P}(\mathbb{N}^{n} )$ be c.c.b$.$ and let $T\subseteq
\mathbb{N}^{n} $ be c.e. Then the set $S=\{x\in \mathbb{N}^{k}
\mid \Phi (x)\subseteq T\}$ is c.e. \qed
\end{enumerate}
\end{prop}

\noindent A function $F:\mathbb{N}^{k} \rightarrow \mathbb{Q}$ is called
\textbf{computable} if there exist computable functions
$a,b,c:\mathbb{N}^{k}\rightarrow \mathbb{N}$ such that
$$F(x)=(-1)^{c(x)}\frac{a(x)}{b(x)+1}$$ for each $x\in
\mathbb{N}^{k} $. A number $x\in \mathbb{R}$ is said to be
\textbf{computable} if there exists a computable function
$g:\mathbb{N}\rightarrow \mathbb{Q}$ such that $|x-g(i)|<2^{-i}$
for each $i\in \mathbb{N}$ \cite{turing}.

By a \textbf{computable} function $\mathbb{N} ^{k} \rightarrow
\mathbb{R}$ we mean a function $f:\mathbb{N} ^{k} \rightarrow
\mathbb{R}$ for which there exists a computable function
$F:\mathbb{N}^{k+1}\rightarrow \mathbb{Q}$ such that
$$|f(x)-F(x,i)|<2^{-i}$$ for all $x\in \mathbb{N}^{k}$ and $i\in
\mathbb{N}$.

In the following proposition we state some  facts about computable
functions $\mathbb{N}^{k} \rightarrow \mathbb{R}.$
\begin{prop} \label{NuR}\hfill
\begin{enumerate}%[\em(1)]

\item If $f,g:\mathbb{N}^{k} \rightarrow \mathbb{R}$ are
computable, then $f+g,f-g,\sqrt{|f|}:\mathbb{N}^{k} \rightarrow
\mathbb{R}$ are computable.

\item If $f:\mathbb{N}^{k} \rightarrow \mathbb{R}$ and
$F:\mathbb{N}^{k+1} \rightarrow \mathbb{R}$ are functions such
that $F$ is computable and $|f(x)-F(x,i)|<2^{-i}$ for each $x\in
\mathbb{N}^{k}$ and $i\in \mathbb{N}$, then $f$ is computable.

\item If $f,g:\mathbb{N}^{k} \rightarrow \mathbb{R}$ are
computable functions, then the set $\{x\in \mathbb{N}^{k} \mid
f(x)>g(x)\}$ is c.e.

\item  If $f:\mathbb{N}^{k}\rightarrow \mathbb{R}$ is a computable
function and $\Phi  :\mathbb{N}^{n} \rightarrow
\mathcal{P}(\mathbb{N}^{k} )$ is a c.c.b.\ function such that
$\Phi (x)\neq\emptyset $ for each $x\in \mathbb{N}^{n}$, then the
functions $g,h:\mathbb{N}^{n}\rightarrow \mathbb{R}$ defined by
$$g(x)=\max_{y\in  \Phi  (x)}f(y),\mbox{ }h(x)=\min_{y\in  \Phi  (x)}f(y)$$  are computable.
\qed

\end{enumerate}
\end{prop}

\subsection{Computable metric spaces}

A tuple $(X,d,\alpha )$ is said to be a \textbf{computable metric
space} if $(X,d)$ is a metric space and $\alpha :\mathbb{N}
\rightarrow X$ is a sequence dense in $(X,d)$ (i.e.\ a sequence
whose range is dense in $(X,d)$) such that the function
$\mathbb{N} ^{2}\rightarrow \mathbb{R} $, $$(i,j)\mapsto
d(\alpha_{i} ,\alpha_{j} )$$ is computable (we use notation
$\alpha =(\alpha _{i} )$).

If  $(X,d,\alpha )$ is a computable metric space, then a sequence
$(x_{i} )$ in $X$ is said to be \textbf{computable} in
$(X,d,\alpha )$ if there exists a computable function
$F:\mathbb{N} ^{2}\rightarrow \mathbb{N} $ such that $$d(x_{i}
,\alpha _{F(i,k)})<2^{-k}$$ for all $i,k\in \mathbb{N} $. A point
$a \in X$ is said to be \textbf{computable} in $(X,d,\alpha )$ if
there exists a computable function $f:\mathbb{N}\rightarrow
\mathbb{N}$ such that $d(a,\alpha _{f(k)})<2^{-k}$ for each $k\in
\mathbb{N}$.

The points   $\alpha _{0}$, $\alpha _{1} $, $\dots$  are called
\textbf{rational points}. If $i\in \mathbb{N}$ and $q\in
\mathbb{Q}$, $q>0$, then we say that  $B(\alpha _{i} ,q)$ is an
(open) \textbf{rational ball} and $\widehat{B}(\alpha _{i} ,q)$ is
a \textbf{closed rational ball}. Here, for $x\in X$ and $r>0$, we
denote by $B(x,r)$ the open ball of radius $r$ centered at $x$ and
by $\widehat{B}(x,r)$ the corresponding closed ball, i.e$.$
$B(x,r)=\{y\in X\mid d(x,y)<r\},$ $\widehat{B}(x,r)=\{y\in X\mid
d(x,y)\leq r\}$.

If $B_{1} ,\dots ,B_{n} $, $n\geq 1$, are open rational balls,
then the union $B_{1}\cup \dots \cup B_{n} $
 will be called a \textbf{rational open set}.

\begin{exa} \label{ex1} If $\alpha :\mathbb{N} \rightarrow
\mathbb{R}^{n}  $ is a computable function (in the sense that the
component functions of $\alpha $ are computable) whose image is
dense in $\mathbb{R}^{n} $ and $d$ is the Euclidean metric on
$\mathbb{R}^{n} $, then $(\mathbb{R}^{n}  , d, \alpha )$ is a
computable metric space (Proposition \ref{NuR}). A sequence
$(x_{i} )$ is computable in this computable metric space if and
only if $(x_{i} )$ is a computable sequence in $\mathbb{R}^{n} $
and $(x_{1} ,\dots ,x_{n} )\in \mathbb{R}^{n} $ is a computable
point in this space if and only if $x_{1} $, \dots, $x_{n} $ are
computable numbers.
\end{exa}

\subsection{Effective enumerations}

Let $(X,d,\alpha )$ be a computable metric space. Let
$q:\mathbb{N}\rightarrow \mathbb{Q}$ be some fixed computable
function whose image is $\mathbb{Q}\cap \langle 0,\infty\rangle $
and let $\tau_{1}   ,\tau_{2}   :\mathbb{N}\rightarrow \mathbb{N}$
be some fixed computable functions such that $\{(\tau _{1} (i),
\tau _{2} (i))\mid i\in \mathbb{N}\}=\mathbb{N}^{2}.$ Let
$(\lambda _{i} )_{i\in \mathbb{N}}$ be the sequence of points in
$X$ defined by $\lambda _{i} =\alpha _{\tau _{1} (i)}$ and let
$(\rho _{i} )_{i\in \mathbb{N}}$ be the sequence of rational
numbers defined by $\rho _{i} =q_{\tau _{2} (i)}$. For $i\in
\mathbb{N}$ we define
$$I_{i}=B(\lambda_{i}   ,\rho_{i}  ),~\widehat{I}_{i}=\widehat{B}(\lambda_{i}   ,\rho_{i}  ).$$ The sequences $(I_{i} )$ and $(\widehat{I}_{i})$ represent  effective
enumerations of all open rational balls and all closed rational
balls.

Now we want to define some effective enumeration of all rational
open sets.

Let $\sigma :\mathbb{N}^{2}\rightarrow \mathbb{N}$ and
$\eta:\mathbb{N}\rightarrow \mathbb{N}$ be some fixed computable
functions with the following property: $\{(\sigma (j,0),\dots
,\sigma (j,\eta(j)))\mid j\in \mathbb{N}\}$ is the set of all
finite sequences in $\mathbb{N}$ (excluding the empty sequence),
i.e. the set $\{(a_{0} ,\dots ,a_{n} )\mid n\in  \mathbb{N},~a_{0}
,\dots ,a_{n} \in \mathbb{N}\}.$ Such functions, for instance, can
be defined using the Cantor pairing function. We use the following
notation: $(j)_{i}$ instead of $\sigma (j,i)$ and $\overline{j}$
instead of $\eta(j).$  Hence $$\{((j)_{0} ,\dots
,(j)_{\overline{j}})\mid j\in \mathbb{N}\}$$ is the set of all
finite sequences in $\mathbb{N}.$ For $j\in \mathbb{N}$ let $[j]$
be defined by
\begin{equation}\label{p2-eq}
[j]=\{(j)_{i} \mid 0\leq i\leq \overline{j}\}.
\end{equation}
Note that the function $\mathbb{N}\rightarrow
\mathcal{P}(\mathbb{N})$, $j\mapsto [j]$, is c.c.b.

For $j\in \mathbb{N}$ we define
$$J_{j}=\bigcup_{i\in [j]}I_{i}.$$
Then  $(J_{j}) $ is an effective enumeration of all rational open
sets.

For $i\in \mathbb{N}$ we define $\Lambda _{i}=\alpha ([i])$, hence
$$\Lambda_{i}=\{\alpha _{(i)_{0} },\dots
,\alpha_{(i)_{\overline{i}}}\}.$$ The sequence $(\Lambda _{i} )$
is an effective enumeration of all finite nonempty sets of
rational points.

\subsection{Formal diameter and formal disjointness}

In Euclidean space $\mathbb{R}^{n}$ we can effectively calculate
the diameter of the finite union of rational balls. However, in a
general computable metric space the function
$\mathbb{N}\rightarrow \mathbb{R}$, $j\mapsto \diam(J_{j} )$, need
not be computable. For that reason we are going to use the notion
of the formal diameter. Let $(X,d)$ be a metric space and
$x_{0},\dots ,x_{k} \in X,$ $r_{0} ,\dots ,r_{k} \in
\mathbb{R}_{+}.$ The \textbf{formal diameter} associated to the
finite sequence $(x_{0},r_{0} ),\dots ,(x_{k},r_{k} ) $ is the
number $D\in \mathbb{R}$ defined by
$$D=\max_{0\leq v,w\leq k }d(x_{v} ,x_{w} )+2\max_{0\leq v\leq
k}r_{v}.$$ It follows from this definition that $\diam(B(x_{0}
,r_{0} )\cup \dots \cup B(x_{k} ,r_{k} ))\leq D$.

Let $(X,d,\alpha )$ be a computable metric space. We define the
function $\fdiam:\mathbb{N}\rightarrow \mathbb{R}$ in the
following way. For $j\in \mathbb{N}$ the number $\fdiam(j)$ is the
formal diameter associated to the finite sequence
$$\left(\lambda _{(j)_{0}} , \rho _{(j)_{0}}\right),\dots ,\left(\lambda
_{(j)_{\overline{j}}} , \rho  _{(j)_{\overline{j}}}\right).$$
Clearly $\diam(J_{j} )\leq \fdiam(j)$ for each $j\in \mathbb{N}$.

Let $i,j\in \mathbb{N}$. We say that $I_{i} $ and $I_{j} $ are
\textbf{formally disjoint} if $$d(\lambda_{i} ,\lambda_{j}
)>\rho_{i} +\rho_{j} .$$ Note that we define this as a relation
between the numbers $i$ and $j$, not the sets $I_{i} $ and $I_{j}
$ (it is possible that for some $i,i',j,j'$ we have that $I_{i}$
and $I_{j} $ are formally disjoint and $I_{i} =I_{i'}$, $I_{j}
=I_{j'}$, but $I_{i'}$ and $I_{j'}$ are not formally disjoint, for
example such a situation can occur if $d$ is the discrete metric
on $X$). If $I_{i} $ and $I_{j} $ are formally disjoint, then
$I_{i} \cap I_{j} =\emptyset $.

Let $i,j\in \mathbb{N}$. We say that $J_{i} $ and $J_{j} $ are
\textbf{formally disjoint} if $I_{k}$ and $I_{l} $ are formally
disjoint for all $k\in [i]$ and $l\in [j]$. Clearly, if $J_{i} $
and $J_{j} $ are formally disjoint, then $J_{i} \cap J_{j}
=\emptyset $.

\begin{prop} \label{fdiam-FD}\hfill
\begin{enumerate}%[(1)]
\item The function $\fdiam:\mathbb{N}\rightarrow \mathbb{R}$ is
computable.

\item The set $\{(i,j)\in \mathbb{N}^{2}\mid J_{i} $ and $J_{j} $
are formally disjoint$\}$ is c.e.
\end{enumerate}
\end{prop}
\proof This follows from Proposition \ref{p1} and Proposition
\ref{NuR} (for details see \cite{jucs}, Proposition 13 and
Proposition 8). \qed

\subsection{Computable sets} Let $(X,d,\alpha )$ be a computable metric space.
We say that $S$ is a \textbf{computably enumerable closed set} in
$(X,d,\alpha )$ if $S$ is a closed subset of $(X,d)$ and if
$$\{i\in \mathbb{N}\mid S \cap I_{i} \neq\emptyset \}$$ is a
c.e$.$ subset of $\mathbb{N}.$ A closed subset of $(X,d)$ is said
to be \textbf{co-computably enumerable closed set} in $(X,d,\alpha
)$  if there exists a computable function $f:\mathbb{N}\rightarrow
\mathbb{N}$ such that
$$X\setminus S=\bigcup _{i\in \mathbb{N}}I_{f(i)}.$$ It is easy to
see that these definitions do not depend on functions $\tau _{1}
$, $\tau _{2}  $ and $q$. We say that $S$ is a \textbf{computable
closed set} in $(X,d,\alpha )$ if $S$ is  a computably enumerable
closed set and a co-computably enumerable closed set
\cite{br-pr,we}.

We say that $K$ is a \textbf{computable compact set} in
$(X,d,\alpha )$ if $K$ is a compact set in $(X,d)$,  the set
$\{j\in \mathbb{N}\mid K\subseteq J_{j} \}$  is c.e. and $K$ is a
c.e. closed set in $(X,d,\alpha )$.

\subsection{Hausdorff metric} Let $(X,d,\alpha )$ be a computable metric space. Let
$A,B\subseteq X$ and $\varepsilon >0$. We write
$A\prec_{\varepsilon }B$ if for each $a\in A$ there exists $b\in
B$ such that $d(a,b)<\varepsilon $. We write
$$A\approx_{\varepsilon }B$$ if $A\prec_{\varepsilon }B$ and $B\prec_{\varepsilon }A$.  Note that
$A\prec_{\varepsilon }B$ and $B\prec_{\varepsilon '}C$ imply
$A\prec_{\varepsilon +\varepsilon '}C$. Also note that
$A\prec_{\varepsilon }B$ and $A'\prec_{\varepsilon}B'$ imply
$A\cup A'\prec_{\varepsilon } B\cup B'$.

Let $\mathcal{H}$ be the set of all nonempty compact subsets of
$(X,d)$. The \textbf{Hausdorff metric} on $\mathcal{H}$ is the
metric $\varrho $ on $\mathcal{H}$ defined by
$$\varrho (A,B)=\inf\{\varepsilon
>0\mid A\approx_{\varepsilon }B\}.$$ If $A,B\in \mathcal{H}$ and $\varepsilon >0$, then
$$A\prec_{\varepsilon }B~\Leftrightarrow~d(a,B)<\varepsilon \mbox{
for each }a\in A~\Leftrightarrow~\max_{a\in A}d(a,B)<\varepsilon
$$ (the last equivalence follows from the fact that the continuous
function $X\rightarrow \mathbb{R}$, $x\mapsto d(x,B)$ attains its
maximum on  $A$). Therefore, for all $A,B\in \mathcal{H}$ we have
\begin{equation}\label{Hausd-metr}
\varrho (A,B)=\max\left\{\max_{a\in A}d(a,B),\max_{b\in
B}d(b,A)\right\}.
\end{equation}
 Also
note that for $A,B\in \mathcal{H}$ and $\varepsilon
>0$ we have
$\varrho (A,B)<\varepsilon$ if and only if $A\approx_{\varepsilon
}B$.

\begin{prop} Let $(X,d,\alpha )$ be a computable metric space. The
function $\mathbb{N}^{2}\rightarrow \mathbb{R}$, $(i,j)\mapsto
\varrho (\Lambda _{i} ,\Lambda _{j} )$ is computable.
\end{prop}
\proof This follows from (\ref{Hausd-metr}) and Proposition
\ref{NuR}(4). \qed

It is easy to conclude that the sequence $(\Lambda _{i} )$ is
dense in the metric space $(\mathcal{H},\varrho )$. Therefore, the
triple $(\mathcal{H},\varrho,\Lambda )$ is a computable metric
space. The following proposition says that computable points in
this space are exactly nonempty computable compact sets in
$(X,d,\alpha )$ (see also Theorem 4.12. in \cite{br-pr}).

\begin{prop} \label{comp-point-in-hiperspace} Let $(X,d,\alpha )$ be a computable metric space and
let $K$ be a nonempty compact set in $(X,d)$. Then $K$ is a
computable compact set in $(X,d,\alpha )$ if and only if there
exists a computable function $f:\mathbb{N}\rightarrow \mathbb{N}$
such that
\begin{equation}\label{comp-point-in-hiperspace-1}
\varrho (K,\Lambda _{f(k)})<2^{-k}
\end{equation}
 for each $k\in \mathbb{N}$.
\end{prop}
\proof Suppose $K$ is a computable compact set. Then the set
$\{i\in \mathbb{N}\mid I_{i} \cap K\neq\emptyset \}$ is c.e.\ and
we conclude from Proposition \ref{p1}(4) that the set $\{j\in
\mathbb{N}\mid I_{i} \cap K\neq\emptyset $ for each $i\in [j]\}$
is c.e. Similarly, using also Proposition \ref{NuR}(3), we get
that the set $$\{(j,k)\in \mathbb{N}^{2}\mid \rho _{i}
<2^{-k}\mbox{ for each }i\in [j]\}$$ is c.e. Therefore the set
$$\Omega =\{(j,k)\in \mathbb{N}^{2}\mid K\subseteq J_{j} , I_{i}
\cap K\neq\emptyset\mbox{ and }\rho _{i} <2^{-k}\mbox{ for each
}i\in [j]\}$$ is c.e. Since for each $k\in \mathbb{N}$ there exist
$j\in \mathbb{N}$ such that $(j,k)\in \Omega $, there exists a
computable function $\varphi :\mathbb{N}\rightarrow \mathbb{N}$
such that $(\varphi (k),k)\in \Omega $ for each $k\in \mathbb{N}$.
Since the function $k\mapsto [k]$ is c.c.b.\ and each nonempty
subset of $\mathbb{N}$ is equal to $[k]$ for some $k\in
\mathbb{N}$, Proposition \ref{p1} implies that there exists a
computable function $f:\mathbb{N}\rightarrow \mathbb{N}$ such that
$[f(k)]=\{\tau _{1} (i)\mid i\in [\varphi (k)]\}$ for each $k\in
\mathbb{N}$. Then $$\Lambda _{f(k)}=\{\lambda _{i}\mid i\in
[\varphi (k)]\}$$ and it follows that $\varrho(K,\Lambda
_{f(k)})<2^{-k}$.

Conversely, suppose that (\ref{comp-point-in-hiperspace-1}) holds
for some computable function $f:\mathbb{N}\rightarrow \mathbb{N}$.
We want to prove that $K$ is a computable compact set.

If $x,y\in X$ and $r,s>0$, we will say that $(y,s)$ is formally
contained in $(x,r)$ and write $(y,s)\subseteq _{F}(x,r)$ if
$d(x,y)+s<r$. Note that $(y,s)\subseteq _{F}(x,r)$ implies
$B(y,s)\subseteq B(x,r)$.

Suppose that $K\subseteq J_{j} $ for some $j\in \mathbb{N}$. Using
compactness of $K$ it is not hard to conclude (see Lemma 19 in
\cite{jucs}) that there exists $\mu
>0$ with property  that for each $x\in K$ there exists $i\in [j]$ such that
$(x,2\mu )\subseteq _{F}(\lambda _{i} ,\rho _{i} )$. Choose $k\in
\mathbb{N}$ so that $2^{-k}<\mu $. Let $l\in [f(k)]$. Since
(\ref{comp-point-in-hiperspace-1}) holds, there exists $x\in K$
such that $d(\alpha _{l} ,x)<\mu $. On the other hand, there
exists $i\in [j]$ such that $(x,2\mu )\subseteq _{F}(\lambda _{i}
,\rho _{i} )$. It follows  $(\alpha _{l} ,\mu )\subseteq
_{F}(\lambda _{i} ,\rho _{i} )$. Hence
\begin{equation}\label{comp-point-in-hiperspace-2}
\left(\forall l\in [f(k)]\right)\mbox{ } \left(\exists i\in
[j]\right) \mbox{ }(\alpha _{l} ,2^{-k} )\subseteq _{F}(\lambda
_{i} ,\rho _{i} ).
\end{equation}
However, if $j,k\in \mathbb{N}$ are such that
(\ref{comp-point-in-hiperspace-2}) holds, then we have $K\subseteq
J_{j} $. Namely, for each $x\in K$ by
(\ref{comp-point-in-hiperspace-1}) there exists $l\in [f(k)]$ such
that $d(x,\alpha _{l} )<2^{-k}$ and, by
(\ref{comp-point-in-hiperspace-2}), there exists $i\in [j]$ such
that $d(\alpha _{l} ,\lambda _{i})+ 2^{-k} <\rho _{i} $ from which
we get $d(x,\lambda _{i})< \rho _{i} $. Hence $x\in J_{j} $.

Therefore $K\subseteq J_{j} $ if and only if there exists $k\in
\mathbb{N}$ such that (\ref{comp-point-in-hiperspace-2}) holds.
Since the set of all $(j,k)\in \mathbb{N}^{2}$ such that
(\ref{comp-point-in-hiperspace-2}) holds is c.e.\ (which follows
from Proposition \ref{NuR} and Proposition \ref{p1}), the set
$\{j\in \mathbb{N}\mid K\subseteq J_{j} \}$ is c.e.

Let $i\in \mathbb{N}$. Then using
(\ref{comp-point-in-hiperspace-1}) we conclude that $$K\cap I_{i}
\neq\emptyset \Leftrightarrow (\exists x\in K)\mbox{ }d(x,\lambda
_{i} )<\rho _{i}\Leftrightarrow (\exists k\in \mathbb{N})\mbox{
}(\exists l\in [f(k)])\mbox{  }d(\alpha _{l} ,\lambda _{i}
)+2^{-k}<\rho _{i}$$ and it follows from Proposition \ref{NuR}
that $K$ is a c.e.\ closed set.  \qed

\section{Semi-computable compact sets} \label{sect-3}

Let $(X,d,\alpha )$ be a computable metric space. We say that $K$
is a \textbf{semi-computable compact set} in $(X,d,\alpha )$ if
$K$ is a compact set in $(X,d)$ and if the set $\{j\in
\mathbb{N}\mid K\subseteq J_{j} \}$  is c.e. Hence $K$ is a
computable compact set if and only if $K$ is a semi-computable
compact set and $K$ is a c.e.\ closed set.

\begin{prop}
Let $(X,d,\alpha )$ be a computable metric space. If $S$ is a
semi-computable compact set in $(X,d,\alpha )$, then $S$ is a
co-c.e.\ closed set in $(X,d,\alpha )$.
\end{prop}
\proof Let $L$ be the set of all $l\in \mathbb{N}$ for which there
exists $j\in \mathbb{N}$ such that $S\subseteq J_{j} $ and such
that $I_{l} $ is formally disjoint with $I_{i} $ for each $i\in
[j]$. Since the set $\{(l,i)\in \mathbb{N}^{2}\mid I_{l} $ and
$I_{i} $ are formally disjoint$\}$ is c.e.\ by Proposition
\ref{NuR}, $L$ is c.e. Clearly $\bigcup_{l\in L}I_{l} \subseteq
X\setminus S$.

On the other hand, let $x\in X\setminus S$. Then, since $S$ is
compact, there exists $\mu >0$ such that $d(x,s)>4\mu $ for each
$s\in S$. Compactness of $S$ furthermore implies that there exists
$j\in \mathbb{N}$ such that $S\subseteq J_{j} $ and such that
$\rho _{i} <\mu $ and $I_{i} \cap S\neq \emptyset $ for each $i\in
[j]$. Choose $l\in \mathbb{N}$ so that $x\in I_{l} $ and $\rho
_{l} <\mu $. It is straightforward to check that $I_{l} $ and
$I_{i} $ are formally disjoint for each $i\in [j]$. Hence $l\in L$
and this proves that $X\setminus S\subseteq \bigcup_{l\in
L}I_{l}$, hence $X\setminus S= \bigcup_{l\in L}I_{l}$. \qed

The preceding proposition implies that each computable compact set
is computable closed set.

 A computable metric space $(X,d,\alpha )$
is \textbf{locally computable}  \cite{brat} if for each compact
set $A$ in $(X,d)$ there exists a computable compact set $K$ in
$(X,d,\alpha )$ such that $A\subseteq K$. For example, the
computable metric space $(\mathbb{R}^{n} ,d,\alpha )$ from Example
\ref{ex1} is locally computable (this can be deduced from
Proposition \ref{comp-point-in-hiperspace}).

\begin{prop} \label{coce-semi}
Let $(X,d,\alpha )$ be a computable metric space which is locally
computable. Let $S$ be a co-c.e.\ closed set in $(X,d,\alpha )$
which is compact. Then $S$ is a semi-computable compact set in
$(X,d,\alpha )$.
\end{prop}
\proof Since $S$ is co-c.e., there exists a computable function
$f:\mathbb{N}\rightarrow \mathbb{N}$  such that $$X\setminus
S=\bigcup_{n\in \mathbb{N}}J_{f(n)}$$ and $J_{f(n)}\subseteq
J_{f(n+1)}$ for each $n\in \mathbb{N}$. Furthermore, there exists
a computable compact set $K$ such that $S\subseteq K$.

Let $j\in \mathbb{N}$. If $S\subseteq J_{j} $, then $K\setminus
J_{j} $ is a compact subset of $X\setminus S$ and therefore there
exists $n\in \mathbb{N}$ such that $K\setminus J_{j} \subseteq
J_{f(n)}$. We have the following conclusion:
\begin{equation}\label{coce-semi-1}
S\subseteq J_{j} \Leftrightarrow (\exists n\in \mathbb{N})\mbox{
}K\subseteq J_{j} \cup J_{f(n)}.
\end{equation}
By Proposition \ref{p1}(1) the function $\mathbb{N}^{2}\rightarrow
\mathcal{P}(\mathbb{N})$, $(j,n)\mapsto [j]\cup [f(n)]$ is c.c.b.\
and therefore by claim (2) of the same proposition there exists a
computable function $g:\mathbb{N}^{2}\rightarrow \mathbb{N}$ such
that $[j]\cup [f(n)]=[g(j,n)]$ for all $j,n\in \mathbb{N}$. Hence
$J_{j} \cup J_{f(n)}=J_{g(j,n)}$ and the set of all $(j,n)\in
\mathbb{N}^{2}$ such that $K\subseteq J_{g(j,n)}$ is  c.e.\ since
$K$ is semi-computable. Now (\ref{coce-semi-1}) implies that $S$
is semi-computable. \qed

In this paper we will prove that in every computable metric space
$(X,d,\alpha )$ the implication $$S\mbox{ semi-computable compact
set }\Rightarrow S\mbox{ computable compact set}$$ holds for each
compact manifold $S$ with computable boundary. Using this and
Proposition \ref{coce-semi} we conclude that if $(X,d,\alpha )$ is
locally computable, then for every compact manifold $S$ with
computable boundary the following implication holds:
\begin{equation}\label{impl-coce-comp}
S\mbox{ co-c.e.\ closed }\Rightarrow S\mbox{ computable closed}.
\end{equation}

\begin{lem} \label{S-bez-Jj0}
Let $(X,d,\alpha )$ be a computable metric space and let $S$ be a
semi-computable compact set in this space. Let $m\in \mathbb{N}$.
Then $S\setminus J_{m} $ is a semi-computable compact set. \qed
\end{lem}
\proof Let $j\in \mathbb{N}$. Then $S\setminus J_{m} \subseteq
J_{j} $ if and only if $S\subseteq J_{j} \cup J_{m} $ and it
follows  that $S\setminus J_{m} $ is semi-computable (similarly as
in the proof of the previous proposition). \qed

\section{Computability up to a set} \label{sect-4}

Let $(X,d,\alpha )$ be a computable metric space. Let
$A,B\subseteq X$, $A\subseteq B$. We say that $A$ is
\textbf{computable up to} $B$ if there exists a computable
function $f:\mathbb{N}\rightarrow \mathbb{N}$ such that
$$A\prec_{2^{-k}}\Lambda _{f(k)}\mbox{ and }\Lambda
_{f(k)}\prec_{2^{-k}}B$$ for each $k\in \mathbb{N}$. Using
Proposition \ref{comp-point-in-hiperspace} we conclude that for
each  nonempty compact subset $A$ of $(X,d)$ the following
equivalence holds:
$$A\mbox{ is a computable compact set }\Leftrightarrow A\mbox{ is
computable up to }A .$$
\begin{prop} \label{unija-A-B-u-X}
Let $(X,d,\alpha )$ be a computable metric space. Suppose $A,B$
and $S$ are subsets of $X$ such that $A\subseteq S$ and
$B\subseteq S$ and such that $A$ and $B$ are computable up to $S$.
Then $A\cup B$ is computable up to $S$.
\end{prop}
\proof Let $f,g:\mathbb{N}\rightarrow \mathbb{N}$ be computable
functions such that $A\prec_{2^{-k}}\Lambda
_{f(k)}\prec_{2^{-k}}S$ and $B\prec_{2^{-k}}\Lambda
_{g(k)}\prec_{2^{-k}}S$ for each $k\in \mathbb{N}$. Then $$A\cup
B\prec_{2^{-k}}(\Lambda _{f(k)}\cup \Lambda
_{g(k)})\prec_{2^{-k}}S$$ for each $k\in \mathbb{N}$. Let
$h:\mathbb{N}\rightarrow \mathbb{N}$ be a computable function such
that $[h(k)]=[f(k)]\cup [g(k)]$ for each $k\in \mathbb{N}$. Then
$\Lambda _{f(k)}\cup \Lambda _{g(k)}=\alpha ([f(k)]\cup
[g(k)])=\Lambda _{h(k)}$ for each $k\in \mathbb{N}$ and  the claim
follows. \qed

Proposition \ref{unija-A-B-u-X} implies: if $A_{1} ,\dots ,A_{n}
\subseteq S$ and $A_{i} $ is computable up to $S $ for each $i\in
\{1,\dots ,n\}$, then $A_{1} \cup \dots \cup A_{n} $ is computable
up to $S$. Therefore, we have the following statement.

\begin{prop} \label{lokal-global}
Let $(X,d,\alpha )$ be a computable metric space. Let $S$ be a
compact subset of $(X,d)$ and suppose $A_{1} ,\dots ,A_{n} $ are
subsets of $S$ such that $S=A_{1} \cup \dots \cup A_{n} $ and such
that $A_{i} $ is computable up to $S $ for each $i\in \{1,\dots
,n\}$. Then $S$ is a computable compact set. \qed
\end{prop}

\section{Localization of computability}\label{sect-5}

That $S$ is a semi-computable compact set in a computable metric
space means that we can effectively enumerate all rational open
sets which contain $S$. However we don't know, in general, how
close $U$ is to $S$ for a given rational open set $U$ containing
$S$ if $S$ is not computable. On the other hand, if we can, for a
given $\varepsilon
>0$, effectively find a rational open set $U$ which contains $S$
and such that $U\prec_{\varepsilon }S$, then we will have
$S\approx_{\varepsilon }U$ and it will follow from Proposition
\ref{comp-point-in-hiperspace} that $S$ is a computable compact
set.

In order to effectively find, for a given $\varepsilon >0$, such a
rational open set $U$, it would be enough to effectively find
finitely many sets $C_{0} ,\dots ,C_{m} $ whose union contains
$S$, such that the diameter of each of these sets is less than
$\varepsilon $ and such that each of these sets intersects $S$.
Then $C_{0} \cup \dots \cup C_{m}\prec_{\varepsilon }S$ and the
computability of $S$ would follow.

This can be done in certain situations. For example, let us
observe the case when $S$ is an arc. Then for each $\varepsilon
>0$ there exists a finite sequence of rational open sets $C_{0} ,\dots
,C_{m}$ whose union contains $S$ such that $C_{i} \cap C_{j}
=\emptyset$ whenever $|i-j|> 1$ and such that $\diam(C_{i}
)<\varepsilon $ for each $i\in \{0,\dots ,m\}$. And if $S$ is a
semi-computable compact set, a sequence $C_{0} ,\dots ,C_{m} $
with these properties can be found effectively. Still, this is not
enough to conclude that each of these sets intersects $S$.
However, if the arc $S$ has computable endpoints $a$ and $b$, then
the sequence $C_{0} ,\dots ,C_{m} $ can be found so that $a\in
C_{0} $ and $b\in C_{m} $. Now we can conclude that each $C_{i} $
intersects $S$, otherwise $C_{0} \cup \dots \cup C_{i-1}$ and
$C_{i+1}\cup \dots \cup C_{m} $ would be disjoint open sets, each
of them would intersect $S$ and $S$ would be contained in their
union which is impossible since $S$ is connected.

Suppose now that $S$ is homeomorphic to $[0,1]\times [0,1] $. We
can proceed in a similar way as in the case of an arc. For a given
$\varepsilon >0$ we want sets $C_{i,j}$, $0\leq i,j\leq m$ whose
union contains $S$, whose diameters are less then $\varepsilon $
and such that $C_{i,j}$ and  $C_{i',j'}$ are disjoint when
$|i-i'|> 1$ or $|j-j'|> 1$. It turns out that, under some
additional assumptions, we can do this in a satisfactory way
\cite{lmcs}, however when at the end we want to conclude that each
of these sets intersects $S$, the fact that $S$ is connected is
not enough for this conclusion and here we need some deeper
topological facts about the space $[0,1]^{n} $ (see the discussion
on pages 5, 6 and 7 in \cite{lmcs}).

 For $n\geq 1$ and $i\in \{1,\dots ,n\}$ let
\begin{equation}\label{def-A}
A^{n} _{i} =\{(x_{1} ,\dots ,x_{n} )\in [-2,2]^{n} \mid x_{i}
=-2\},
\end{equation}
\begin{equation}\label{def-B}
B^{n} _{i} =\{(x_{1} ,\dots ,x_{n} )\in [-2,2]^{n} \mid x_{i}
=2\}.
\end{equation}
(Here $[-2,2]^{n}$ denotes the set $\{(x_{1} ,\dots ,x_{n} )\mid
x_{1} ,\dots ,x_{n} \in [-2,2]\}$.)
 When the context is
clear, we write $A_{i} $ and $B_{i} $ instead of $A^{n}_{i} $ and
$B^{n}_{i} $.

The following topological fact will be crucial in the proof of our
main result (see \cite{engel}, Theorem 1.8.1, and \cite{lmcs},
Corollary 3.3.).

\begin{thm}\label{glavni} Let $n\geq 1$. Suppose $U_{1} ,\dots ,U_{n} $
and $V_{1} ,\dots ,V_{n} $ are open subsets of $\mathbb{R}^{n} $
such that
$$U_{i} \cap A_{i}=\emptyset ,~V_{i} \cap B_{i}=\emptyset
\mbox{ and }U_{i} \cap V_{i} =\emptyset $$ for all $i\in
\{1,\dots,n\}$. Then $[-2,2]^{n}$ is not contained in the union $
U_{1} \cup \dots \cup U_{n} \cup V_{1} \cup \dots \cup V_{n}  $.
\qed
\end{thm}

Let $X$ be a set, $n\geq 1$ and $m\in \mathbb{N}$. A function
$$C:\mathbb{N}_{m} ^{n}  \rightarrow \mathcal{P}(X)$$  is called
an $n$-\textbf{chain} in $X$ (of length $m$)   if
\begin{equation}\label{sect2-1}
C_{i_{1} ,\dots ,i_{n} }\cap C_{j_{1} ,\dots ,j_{n} }=\emptyset
\end{equation}
for all $(i_{1} ,\dots ,i_{n} ), (j_{1} ,\dots ,j_{n})\in
\mathbb{N}_{m} ^{n}$ such that $|i_{l} -j_{l} |>1$ for some $l\in
\{1,\dots ,n\}$. Here we use $C_{i_{1} ,\dots ,i_{n}}$ to denote
$C(i_{1} ,\dots ,i_{n})$.

For $a,b\in \mathbb{N}^{n} _{m} $, $a=(a_{1} ,\dots ,a_{n} )$,
$b=(b_{1} ,\dots ,b_{n} )$, we define the number $p(a,b)$ by
$$p(a,b)=\max_{1\leq i\leq n}|a_{i} -b_{i} |.$$ With this
notation, we have that $C:\mathbb{N}^{n} _{m} \rightarrow
\mathcal{P}(X)$ is an $n$-chain if and only if $C_{a}\cap
C_{b}=\emptyset $ for all $a,b\in \mathbb{N}^{n} _{m} $ such that
$p(a,b)>1$. If $a\in \mathbb{N}^{n} _{m} $, we say that $C_{a}$ is
a \textbf{link} of the chain $C$.

If $(X,d)$ is a metric space, then we say that an $n-$chain
$C=(C_{i_{1} ,\dots ,i_{n} })_{0\leq i_{1} ,\dots ,i_{n} \leq m}$
in $X$ is \textbf{open} if $C_{i_{1} ,\dots ,i_{n} }$ is an open
set in $(X,d)$ for all $i_{1} ,\dots ,i_{n} \in  \mathbb{N}_{m} $.
We similarly define the notion of a \textbf{compact} $n$-chain in
$(X,d)$.

If $C:\mathbb{N}_{m} ^{n} \rightarrow \mathcal{P}(X)$,
$C=(C_{i_{1} ,\dots ,i_{n} })_{0\leq i_{1} ,\dots ,i_{n} \leq m}$
is a function, then the function
$$(C_{i_{1} ,\dots ,i_{n-1},0 })_{0\leq i_{1} ,\dots ,i_{n-1} \leq
m}$$ is called the \textbf{lower boundary} of $C$. If $C$ is an
$n-$chain in $(X,d)$, then its lower boundary is an $n-1$-chain in
$(X,d)$.

In general, if $A$ is a set and $f:A\rightarrow \mathcal{P}(X)$ a
function, we will denote by $\bigcup f$ the union $\bigcup_{a\in
A}f(a)$ and we will say that $f$ \textbf{covers} $S$, where $S
\subseteq   X$, if $S\subseteq \bigcup f$. If $(X,d)$ is a metric
space, $A$ a nonempty finite set and $f(a)$ a nonempty bounded set
for each $a\in A$, then we define $\mesh(f)$ as the number
$$\mesh(f)=\max_{a\in A}\left(\diam f(a)\right).$$

Let $\varepsilon >0$. An $n$-chain $C$ in a metric space $(X,d)$
is said to be an $\varepsilon -n$-\textbf{chain} if
$\mesh(C)<\varepsilon $.

Let $n\geq 1$. A \textbf{finite} $n-$\textbf{sequence} in
$\mathbb{N}$ is any function of the form $$\{0,\dots ,m\}^{n}
\rightarrow \mathbb{N}.$$ Recall that any finite sequence $i_{0}
,\dots ,i_{m} $ in $\mathbb{N}$ is of the form $(j)_{0} ,\dots,
(j)_{\overline{j}}$ for some $j\in \mathbb{N}$. Let
$\nu:\mathbb{N}^{n} \rightarrow \mathbb{N}$ be some fixed
computable injection and let $\tau _{1}   $ and $\tau _{2}  $ be
the functions from the section \ref{prelim}. We define
$\Sigma:\mathbb{N}^{n+1}\rightarrow \mathbb{N}$ by
$$\Sigma(i,j_{1} ,\dots ,j_{n} )=(\tau _{1}   (i))_{\nu (j_{1} ,\dots ,j_{n}
)}.$$ Then for any finite $n-$sequence $a$ in $\mathbb{N}$ there
exists $i\in \mathbb{N}$ such that $a$ equals the function
$$\{0,\dots ,\tau _{2}  (i)\}^{n} \rightarrow \mathbb{N},$$ $$(j_{1} ,\dots
,j_{n} )\mapsto \Sigma(i,j_{1} ,\dots ,j_{n} ).$$ We will use the
following notation: $\widehat{i}$ instead of $\tau _{2}  (i)$ and,
for $n\geq 2$, $(i)_{j_{1} ,\dots ,j_{n} }$ instead of
$\Sigma(i,j_{1} ,\dots ,j_{n} )$.

Let $(X,d,\alpha )$ be a computable metric space. For $l\in
\mathbb{N}$ let $\mathcal{H}_{l}$ be the finite $n-$sequence of
sets in $X$ defined by
$$\mathcal{H}_{l}=\left(J_{(l)_{j_{1} ,\dots ,j_{n} }}\right)_{0\leq
j_{1} ,\dots ,j_{n} \leq \widehat{l}}$$ (i.e$.$ $\mathcal{H}_{l}$
is the function $\{0,\dots ,\widehat{l}\}^{n} \rightarrow
\mathcal{P}(X)$ which maps $(j_{1} ,\dots ,j_{n} )$ to
$J_{(l)_{j_{1} ,\dots ,j_{n} }}$).

\begin{lem} \label{lem-zeta}
Let $(X,d,\alpha )$ be a computable metric space. Then there exist
computable functions $\zeta ,\zeta ':\mathbb{N}\rightarrow
\mathbb{N}$ such that, for each $l\in \mathbb{N}$, the set
$J_{\zeta (l)}$ is the union of  $\mathcal{H}_{l}$  and $J_{\zeta'
(l)}$ is the union of the lower boundary of $\mathcal{H}_{l}$.
\end{lem}
\proof See Lemma 4.3.\ in \cite{lmcs}. \qed

\begin{prop} \label{H-l-covers}
Let $(X,d,\alpha )$ be a computable metric space and let $S$ be a
semi-computable compact set in this space.
\begin{enumerate}
\item[(i)] The set $\{l\in \mathbb{N}\mid \mathcal{H}_{l} $ covers
$S\}$ is c.e. \item[(ii)] The set $\{l\in \mathbb{N}\mid $ the
lower boundary of $\mathcal{H}_{l} $ covers $S\}$ is c.e.
\end{enumerate}
\end{prop}
\proof Let $\zeta ,\zeta $ be the functions from Lemma
\ref{lem-zeta}. Then $\mathcal{H}_{l} $ covers $S$ if and only if
$S\subseteq J_{\zeta (l)}$ and the lower boundary of
$\mathcal{H}_{l} $ covers $S$ if and only $S\subseteq J_{\zeta
'(l)}$. Claims (i) and (ii) follow. \qed

We say that $\mathcal{H}_{l} $ is a \textbf{formal} $n$-chain if
$J_{(l)_{a}}$ and $J_{(l)_{b}}$ are formally disjoint for all
$a,b\in \mathbb{N}^{n} _{\widehat{l}}$ such that $p(a,b)>1$.

Let the function $\fmesh:\mathbb{N}\rightarrow \mathbb{R}$ be
defined by
$$\fmesh(l)=\max_{0\leq j_{1} ,\dots ,j_{n} \leq \widehat{l}}\fdiam((l)_{j_{1} ,\dots ,j_{n} }).$$
\begin{prop} \label{H-l-formal}\hfill
\begin{enumerate}[label=(\roman*)]
\item The function $\fmesh$ is computable.

\item The set $\{l\in \mathbb{N}\mid \mathcal{H}_{l} $ is a
formal $n$-chain$\}$ is computably enumerable.
\end{enumerate}
\end{prop}
\proof (i) follows from Proposition \ref{fdiam-FD}(1) and
Proposition \ref{NuR}(4).

Let us prove (ii). Let $\Phi :\mathbb{N}\rightarrow
\mathcal{P}(\mathbb{N}^{2n+1})$ be defined by $\Phi (l)=\{(l,v_{1}
,\dots ,v_{n} ,w_{1} ,\dots $ $\dots ,w_{n})\mid 0\leq v_{1}
,\dots ,v_{n} ,w_{1} ,\dots,w_{n}\leq \widetilde{l},$ $|v_{i}
-w_{i} |>1$ for some $i\in \{1,\dots ,n\}\}$. Then $\Phi $ is
c.c.b. Let $\Psi :\mathbb{N}^{2n+1}\rightarrow
\mathcal{P}(\mathbb{N}^{2})$ be defined by $$\Psi (l,v_{1} ,\dots
,v_{n} ,w_{1} ,\dots,w_{n})=\{((l)_{v_{1} ,\dots
,v_{n}},(l)_{w_{1} ,\dots,w_{n}})\}.$$ Clearly, $\Psi$ is c.c.b.
The set $\Omega $ of all $(i,j)\in \mathbb{N}^{2}$ such that
$J_{i} $ and $J_{j} $ are formally disjoint is c.e.\ by
Proposition \ref{fdiam-FD}(2). Finally, the function $\Lambda
:\mathbb{N}\rightarrow \mathcal{P}(\mathbb{N}^{2})$  defined by
$\Phi $ and $\Psi $ as in Proposition \ref{p1}(3) is c.c.b.\ and
we have that $\mathcal{H}_{l} $ is a formal $n-$chain if and only
if $\Lambda (l)\subseteq \Omega $. It follows now from Proposition
\ref{p1}(4) that the set $\{l\in \mathbb{N}\mid \mathcal{H}_{l} $
is a formal $n$-chain$\}$ is c.e. \qed

If $(X,d)$ is a metric space, then for nonempty subsets $S$ and
$T$ of $X$ we denote the number $\inf\{d(x,y)\mid x\in S,~y\in
T\}$ by $d(S,T)$.  The proof of the following lemma is
straightforward.
\begin{lem} \label{K-disj-L-FD}
Let $(X,d)$ be a metric space and let $K$ and $L$ be nonempty
compact sets in this space. Let $r=\frac{d(K,L)}{4}$. Then, if
$x,y\in X$ and $s,t\in \mathbb{R}$ are such that $0<s,t<r$ and
$B(x,s)\cap K\neq\emptyset $ and $B(y,t)\cap L\neq\emptyset $,
then $d(x,y)>s+t$. \qed
\end{lem}

Let us suppose that $S$ is a semi-computable compact set in a
computable metric space $(X,d,\alpha )$. Suppose $S$ is
homeomorphic to $[0,1]\times [0,1]$. To show that $S$ is a
computable compact set, it would be enough to effectively find,
for a given $\varepsilon
>0$, an $\varepsilon -2$-chain which covers $S$ and such that each
of its links intersects $S$. The question whether a chain covers
$S$ or not is semi-decidable since $S$ is a semi-computable
compact set. But the problem is how to ensure that each link of a
chain intersects $S$. In fact, this is not possible in general
since $S$ needs not be computable. The idea we used to tackle this
problem is to find some effective procedure which will, for a
given $\varepsilon
>0$, give an $\varepsilon -2$-chain which covers $S$ in such a way
that we can be sure that certain links (which we can effectively
determine) do intersect $S$. This idea is used in the proof of the
following theorem.

\begin{thm} \label{lok-izr-eukld-skupa}
Let $(X,d,\alpha )$ be a computable metric space. Let $S$ be a
semi-computable compact set in this space and suppose $x\in S$ is
a point which has a neighborhood in $S$ homeomorphic to
$\mathbb{R}^{n} $ for some $n\in \mathbb{N}$. Then there exists a
neighborhood of $x$ in $S$ which is computable up to $S$.
\end{thm}
\proof Let $U$ be a neighborhood of $x$  in $S$ (i.e.\ $x\in
\Int_{S} U$, where $\Int_{S} U$ is the interior of $U$ in $S$) and
$n\in \mathbb{N}$ so that $U$ is homeomorphic to $\mathbb{R}^{n}
$. We may assume that $U$ is open in $S$ (certainly $\Int_{S} U$
is homeomorphic to an open subset of $\mathbb{R}^{n} $ which
implies that some open subset of $\Int_{S} U$ which contains $x$
is homeomorphic to an open ball in $\mathbb{R}^{n} $; however
every open ball in $\mathbb{R}^{n} $ is homeomorphic to
$\mathbb{R}^{n} $). Let $f:\mathbb{R}^{n} \rightarrow U$ be a
homeomorphism. We may assume $f(0)=x$ (otherwise we take the
composition of $f$ and some translation $\mathbb{R}^{n}
\rightarrow \mathbb{R}^{n} $). The set $f(\langle -4,4\rangle ^{n}
)$ is open in $U$ and therefore in $S$ which implies that
$S\setminus f(\langle -4,4\rangle ^{n} )$ is closed in $S$. (Here,
for $a,b\in \mathbb{R}$, we use $\langle a,b\rangle $ to denote
the open interval $\{x\in \mathbb{R}\mid a<x<b\}$.) Hence
$S\setminus f(\langle -4,4\rangle ^{n} )$ is compact and since it
is disjoint with $f([-2,2]^{n} )$, which is clearly compact, there
exists $m_{0} \in \mathbb{N}$ such that the rational open set
$J_{m_{0} }$ satisfies the following:
$$S\setminus f(\langle -4,4\rangle ^{n} )\subseteq J_{m_{0}
}\mbox{ and }J_{m_{0} }\cap f([-2,2]^{n} )=\emptyset .$$ Let
$S'=S\setminus J_{m_{0} }$. Then, by Lemma \ref{S-bez-Jj0}, $S'$
is a semi-computable compact set in $(X,d,\alpha )$ and
\begin{equation}\label{thm-10}
f([-2,2]^{n} )\subseteq S'\subseteq f([-4,4]^{n} ).
\end{equation}
See Figure 1. The areas bounded by the inside, the middle  and the
outside curve are $f([-1,1]^{n}
  )$, $f([-2,2]^{n} )$ and $f([-4,4]^{n} )$ respectively. The grey
  area is $S'$.
\begin{center}
  \includegraphics[height=1.3in,width=1.8in]{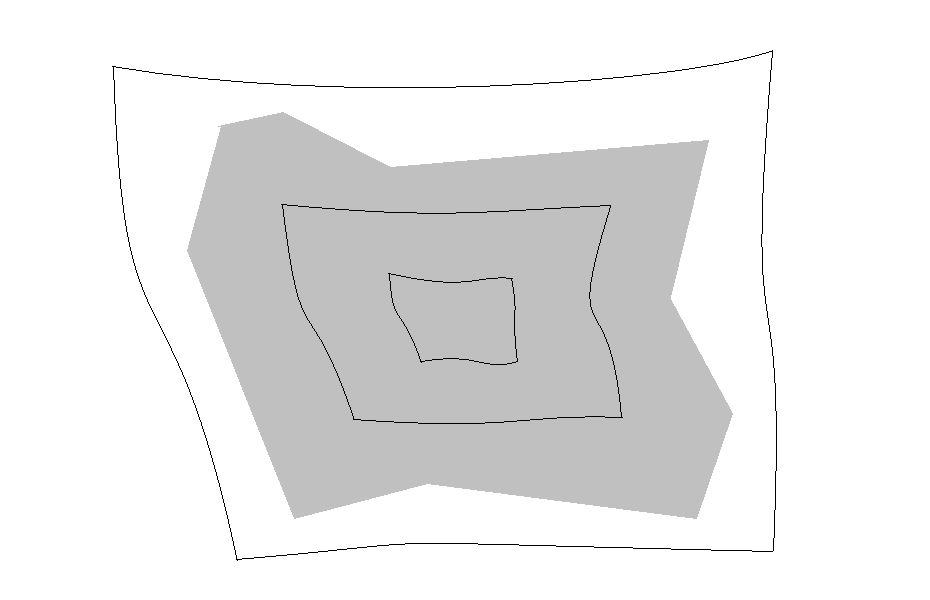}

  \emph{Figure 1.}
\end{center}
For $i\in \{1,\dots ,n\}$ let $$C_{i} =\{(x_{1} ,\dots ,x_{n} )\in
[-4,4]^{n} \mid x_{i} \geq -1\},~~D_{i} =\{(x_{1} ,\dots ,x_{n}
)\in [-4,4]^{n} \mid x_{i} \leq 1\}.$$ Note that for each $i\in
\{1,\dots ,n\}$ the sets $f(A_{i} )$ and $f(C_{i} )$ are compact
and disjoint and the sets $f(B_{i} )$ and $f(D_{i} )$ are compact
and disjoint (recall that $A_{i} $ and $B_{i} $ are defined by
(\ref{def-A}) and (\ref{def-B})). On Figure 2 the blue area is
$f(A_{i}) $ and the green area is $f(C_{i}) $. On Figure 3 the
blue area is $f(B_{i}) $ and the green area is $f(D_{i}) $.

\begin{center}
  \includegraphics[height=1.1in,width=1.5in]{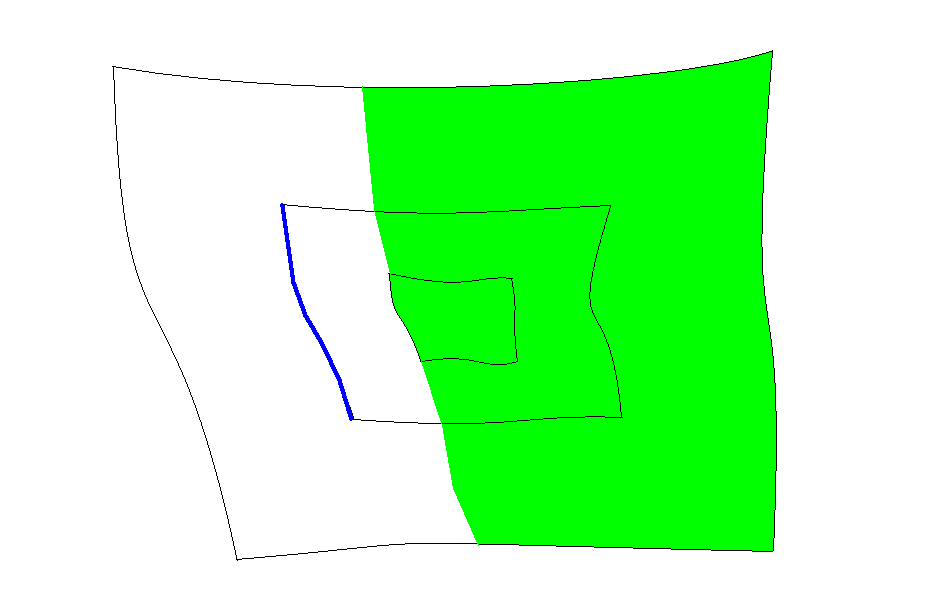}\hspace{50pt}
  \includegraphics[height=1.1in,width=1.5in]{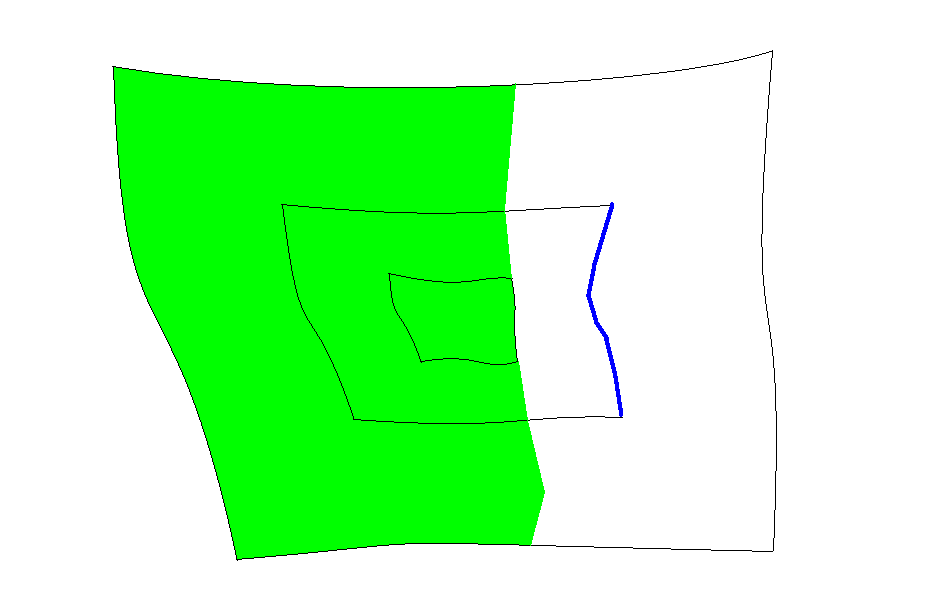}

  \emph{Figure 2.} \hspace{110pt} \emph{Figure 3.}
\end{center}
Also, $\{x\}$ and $f([-4,4]^{n} \setminus \langle -1,1\rangle ^{n}
)$ are compact and disjoint. Choose a positive rational number
$\gamma $ such that
\begin{equation}\label{tm-gama}
\gamma <\frac{d(f(A_{i} )),f(C_{i} ))}{4},\mbox{ }\gamma
<\frac{d(f(B_{i} )),f(D_{i} ))}{4}\mbox{ and }\gamma
<\frac{d(\{x\},f([-4,4]^{n} \setminus \langle -1,1\rangle ^{n}
))}{4}
\end{equation}
 for each $i\in \{1,\dots ,n\}$.

Let $i\in \{1,\dots ,n\}$. Let $\mathcal{U}=\{B(\alpha _{l}
,\gamma )\mid l\in \mathbb{N}\}$. Then $\mathcal{U}$ is an open
cover for $f(A_{i} )$ in $(X,d)$ and therefore there exists a
finite subcover $\mathcal{U}'$ of $\mathcal{U}$ for $f(A_{i} )$.
Let $\mathcal{U}''$ be the subset of $\mathcal{U}'$ consisting of
those members of $\mathcal{U}'$ which intersect $f(A_{i} )$. We
have $$\mathcal{U}''=\{B(\alpha _{l_{0} },\gamma ),\dots ,B(\alpha
_{l_{k} },\gamma )\},$$ where $l_{0} ,\dots ,l_{k} \in
\mathbb{N}$. Choose $\omega \in \mathbb{N}$ such that $\gamma
=q_{\omega }$.  Let $v_{0} ,\dots ,v_{k} \in \mathbb{N}$ be such
that $\tau _{1}  (v_{j} )=l_{j} $ and $\tau _{2}  (v_{j} )=\omega
$ for each $j\in \{0,\dots ,k \}$. Finally, choose  $a_{i} \in
\mathbb{N}$ so that
$$[a_{i} ]=\{v_{0} ,\dots ,v_{k}  \}.$$
In this way we have defined the numbers $a_{1} ,\dots ,a_{n} $.
For each $i\in \{1,\dots ,n\}$ clearly $f(A_{i} )\subseteq
J_{a_{i} }$ and if $j\in \mathbb{N}$ is such that $I_{j} \cap
f(C_{i} )\neq \emptyset $ and $\rho_{j}<\gamma $, then by
(\ref{tm-gama}) and  Lemma \ref{K-disj-L-FD} we have that $I_{j} $
and $J_{a_{i} }$ are formally disjoint.

In the same way we define numbers $b_{1} ,\dots ,b_{n} $ with the
following property:  if $i\in \{1,\dots ,n\}$, then  $f(B_{i}
)\subseteq J_{b_{i} }$ and if $j\in \mathbb{N}$ is such that
$I_{j} \cap f(D_{i} )\neq \emptyset $ and $\rho_{j}<\gamma $,
then $I_{j} $ and $J_{b_{i} }$ are formally disjoint.

Similarly, there exists $x' \in \mathbb{N}$ with the property that
$x\in J_{x' }$ and if $j\in \mathbb{N}$ is such that $I_{j} \cap
f([-4,4]^{n} \setminus \langle -1,1\rangle ^{n} )\neq \emptyset$
and $\rho_{j}<\gamma $, then $I_{j} $ and $J_{x'}$ are
formally disjoint.

For $m\in \mathbb{N}$ let $E^{m} :\mathbb{N}_{8m+7} ^{n}
\rightarrow \mathcal{P}([-4,4]^{n}  )$ be defined by
$$E^{m}_{i_{1} ,\dots ,i_{n} }=\left[-4+\frac{i_{1}
}{m+1},-4+\frac{i_{1} +1}{m+1}\right]\times \dots \times
\left[-4+\frac{i_{n} }{m+1},-4+\frac{i_{n} +1}{m+1}\right].$$ Then
$E^{m} $ is a compact $n$-chain in $[-4,4]^{n} $ which covers
$[-4,4]^{n} $. Let $F^{m} :\mathbb{N}_{8m+7} ^{n} \rightarrow
\mathcal{P}(X  )$ be defined by $F^{m} (v)=f(E^{m} (v))$,
$v\in\mathbb{N}_{8m+7} ^{n} $. Then $F^{m} $ is a compact
$n$-chain in $(X,d)$ which covers $f([-4,4]^{n}) $.

\begin{center}
  \includegraphics[height=1.1in,width=1.7in]{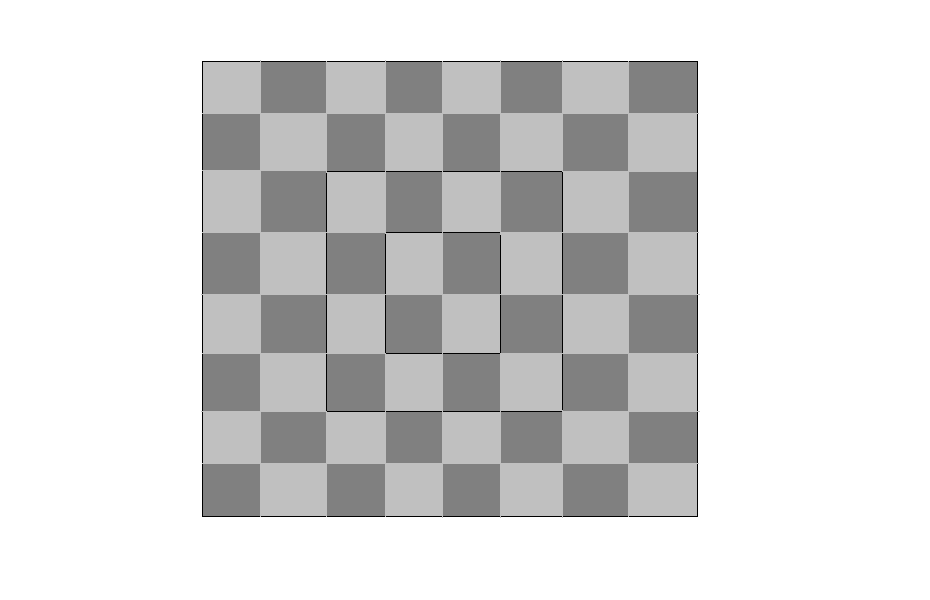}\hspace{35pt}
  \includegraphics[height=1.1in,width=1.5in]{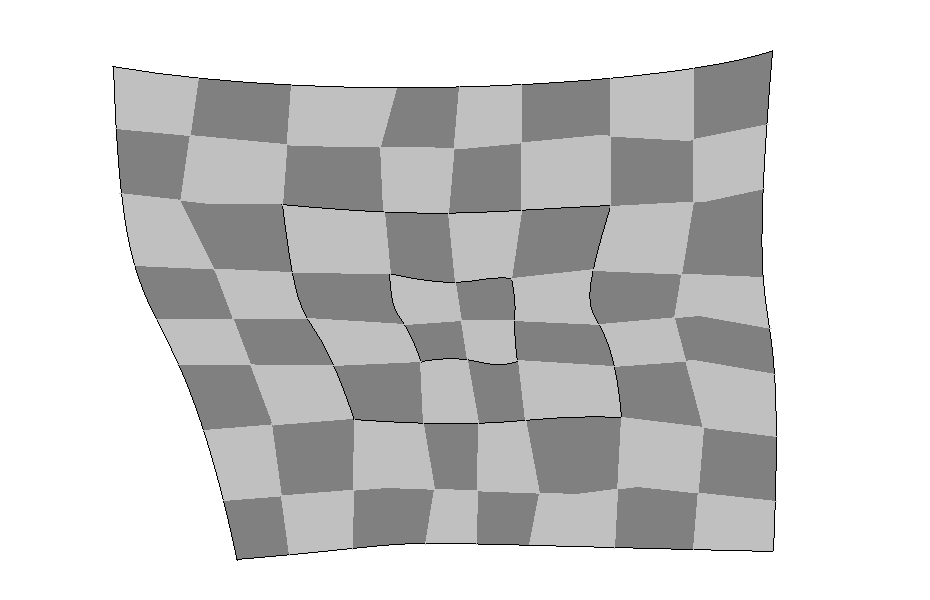}

  \emph{Figure 4.} Chain $E^{m} $ \hspace{70pt} \emph{Figure 5.} Chain $F^{m}
  $
\end{center}

Let $\varepsilon >0$. Since the restriction of $f$ to $[-4,4]^{n}
$ is uniformly continuous, there exists $m\in \mathbb{N}$ such
that $\mesh (F^{m} )<\frac{\varepsilon }{6}$. Let $\delta $ be a
positive rational number such that $\delta <\gamma $, $\delta
<\frac{\varepsilon }{6}$ and  $$\delta <\frac{d(F^{m} (v),F^{m}
(w))}{4}$$ for all $v,w\in \mathbb{N}^{n} _{8m+7}$ such that
$p(v,w)>1$.

Now we define numbers $j_{v}$ for $v\in \mathbb{N}_{8m+7}$ in the
following way. Let $v\in \mathbb{N}_{8m+7}^{n} $. As before, we
conclude that there exist finitely many numbers $i_{0} ,\dots
,i_{k}$ such that $F^{m} (v)$ is contained in the union of the
balls $B(\alpha _{i_{0} },\delta ),\dots, B(\alpha _{i_{k}
},\delta )$ and each of these balls intersects $F^{m} (v)$. Let
$p\in \mathbb{N}$ be such that $q_{p}=\delta $. Now, let $w_{0}
,\dots ,w_{k}$ be such that $\tau _{1}  (w_{l} )=i_{l} $, $\tau
_{2} (w_{l} )=p$ for each $l\in \{0,\dots ,k \}$. Finally, choose
$j_{v}\in \mathbb{N}$ so that $[j_{v}]=\{w_{0} ,\dots ,w_{m} \}$.

We conclude readily that $\fdiam (j_{v})<\varepsilon $ for each
$v\in \mathbb{N}_{8m+7}^{n}$ and it also follows, using Lemma
\ref{K-disj-L-FD}, that $J_{j_{v}}$ and $J_{j_{w}}$ are formally
disjoint for all $v,w\in \mathbb{N}_{8m+7}^{n}$ such that
$p(v,w)>1$. Furthermore, let $v\in \mathbb{N}_{8m+7}^{n}$,
$v=(v_{1} ,\dots ,v_{n} )$ and let $i\in \{1,\dots ,n\}$.  If
$v_{i} \geq 3m+3$, then $F^{m} (v)\subseteq f(C_{i} )$ and by the
construction of $j_{v}$ and $a_{i} $ we have that $J_{j_{v}}$ and
$J_{a_{i} }$ are formally disjoint. Similarly, if $v_{i} \leq
5m+4$, then $F^{m} (v)\subseteq f(D_{i} )$ and we get that
$J_{j_{v}}$ and $J_{b_{i} }$ are formally disjoint.  Also note
that $v_{i} <3m+3$ or $v_{i}
>5m+4$ implies that $J_{j_{v}}$ and $J_{x'}$ are formally
disjoint. Note that $S'$ (in fact $f([-4,4]^{n} )$ is contained in
the union of the sets $J_{j_{v}}$, $v\in \mathbb{N}_{8m+7}^{n}$.

Choose $l\in \mathbb{N}$ such that $\widehat{l}=8m+7$ and such
that $j_{v}=(l)_{v}$ for each $v\in \mathbb{N}_{8m+7}^{n} $. Put
$e=3m+3$, $h=5m+4$. The preceding construction can be made for
$\varepsilon =2^{-k}$ for each $k\in \mathbb{N}$. We have the
following conclusions:
\begin{enumerate}%[(1)]
\item $\mathcal{H}_{l} $ covers $S'$;

\item $\mathcal{H}_{l} $ is a formal $n$-chain;

\item $J_{(l)_{v_{1} ,\dots ,v_{n} }}$ and $J_{a_{i} }$ are
formally disjoint if $1\leq i\leq n$, $0\leq v_{1} ,\dots ,v_{n}
\leq \widehat{l}$ and  $v_{i} \geq e$;

\item $J_{(l)_{v_{1} ,\dots ,v_{n} }}$ and $J_{b_{i} }$ are
formally disjoint if $1\leq i\leq n$, $0\leq v_{1} ,\dots ,v_{n}
\leq \widehat{l}$ and  $v_{i} \leq  h$;

\item $J_{(l)_{v_{1} ,\dots ,v_{n} }}$ and $J_{x' }$ are formally
disjoint if $1\leq i\leq n$, $0\leq v_{1} ,\dots ,v_{n} \leq
\widehat{l}$ and  ($v_{i}  <e$ or $v_{i} >h$);

\item $\fmesh (l)<2^{-k}$.

\end{enumerate}

These conditions are semi-decidable, i.e.\ the set $\Omega $ of
all $(k,l,e,h)\in \mathbb{N}^{4}$ such that (1)--(6) hold is c.e.\
(Proposition \ref{H-l-covers}, Proposition \ref{H-l-formal} and
Proposition \ref{NuR}(3) imply that conditions (1), (2) and (6)
are semi-decidable; that (3)--(5) are semi-decidable can be seen
as in the proof of Proposition \ref{H-l-formal}). Since for each
$k\in \mathbb{N}$ there exist $l,e,h\in \mathbb{N}$ such that
$(k,l,e,h)\in \Omega $, there exist computable functions
$\tilde{l},\tilde{e},\tilde{h}:\mathbb{N}\rightarrow \mathbb{N}$
such that $(k,\tilde{l}(k),\tilde{e}(k),\tilde{h}(k))\in \Omega $
for each $k\in \mathbb{N}$. Let $k\in \mathbb{N}$ and let
$l=\tilde{l}(k)$, $e=\tilde{e}(k)$ and $h=\tilde{h}(k)$. Let
$\Gamma$ be the set of all $(v_{1} ,\dots ,v_{n} )\in
\mathbb{N}^{n} _{\widehat{l}}$ such that $e\leq v_{i}\leq h$ for
each $i\in \{1,\dots ,n\}$. We claim that
\begin{equation}\label{thm-C1}
J_{(l)_{v}}\cap f([-2,2]^{n} )\neq \emptyset \mbox{ for all }v\in
\Gamma,
\end{equation}
and
\begin{equation}\label{thm-C2}
S'\cap J_{x' }\subseteq \bigcup_{v\in \Gamma }J_{(l)_{v}}.
\end{equation}
Suppose $J_{(l)_{v}}\cap f([-2,2]^{n} )=\emptyset$ for some
$v=(v_{1} ,\dots ,v_{n} )\in \Gamma$.
\begin{center}
  \includegraphics[height=1.1in,width=1.5in]{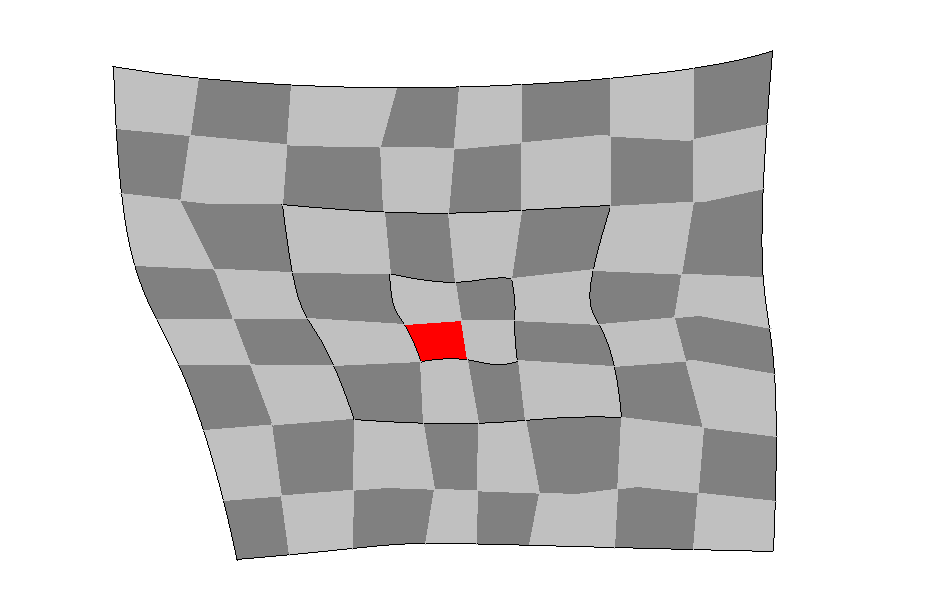}\hspace{20pt}
  \includegraphics[height=1.1in,width=1.5in]{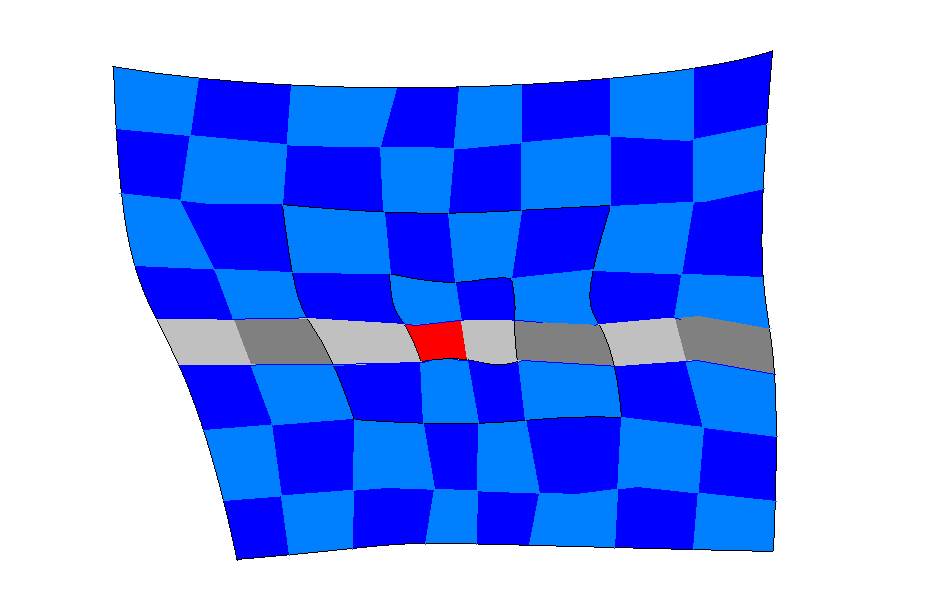}\hspace{20pt}
  \includegraphics[height=1.1in,width=1.5in]{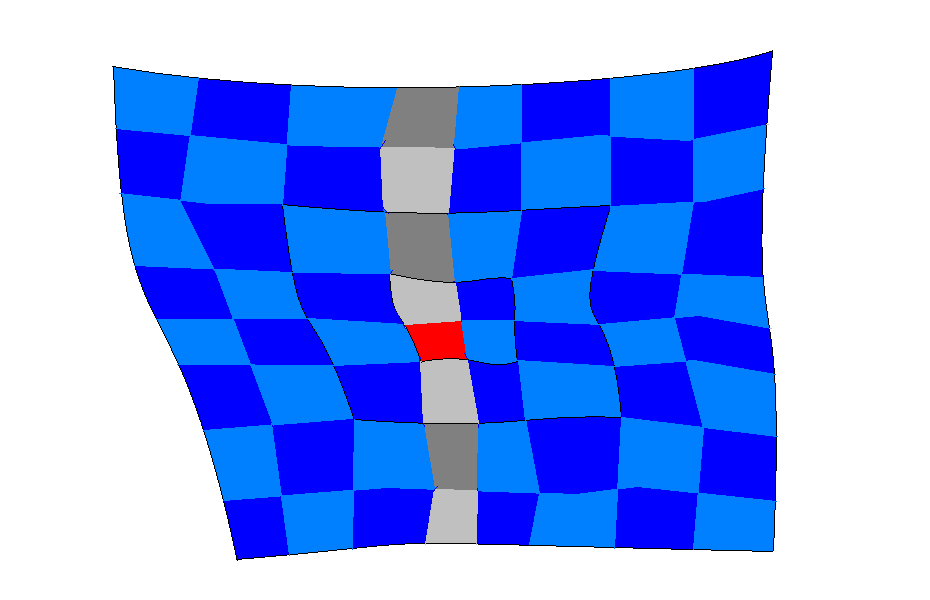}

  \emph{Figure 6.} The red link is $J_{(l)_{v}}$.
\end{center}
 For each $i\in \{1,\dots ,n\}$
we define $U_{i} $ as the union of all $J_{(l)_{w_{1} ,\dots
,w_{n} }}$ such that $(w_{1} ,\dots ,w_{n} )\in \mathbb{N}^{n}
_{\widehat{l}}$ and $w_{i} <v_{i} $ and we define $V_{i} $ as the
union of all $J_{(l)_{w_{1} ,\dots ,w_{n} }}$ such that $(w_{1}
,\dots ,w_{n} )\in \mathbb{N}^{n} _{\widehat{l}}$ and $w_{i}
>v_{i} $. (In Figure 6 the sets $U_{i} $ are represented by the bottom and the left blue
set  and the sets $V_{i} $  by the upper and the right blue set.)
Then $$f([-2,2]^{n} )\subseteq (U_{1} \cup \dots \cup  U_{n} )\cup
(V_{1} \cup \dots \cup V_{n} ).$$ Since $\mathcal{H}_{l}$ is a
chain, $U_{i} \cap V_{i} =\emptyset $ for each $i\in \{1,\dots
,n\}$. Since $v_{i} \leq h$, by property (4) we have $U_{i} \cap
J_{b_{i} }=\emptyset $, hence $U_{i} \cap f(B_{i} )=\emptyset $.
Similarly, $v_{i} \geq e$ implies $V_{i} \cap J_{a_{i} }=\emptyset
$ and this gives $V_{i} \cap f(A_{i} )=\emptyset $. Therefore
$$f^{-1} (U_{i} )\cap B_{i} =\emptyset,\mbox{ }f^{-1} (V_{i} )\cap
A_{i} =\emptyset\mbox{ and }f^{-1} (U_{i} )\cap f^{-1} (V_{i}
)=\emptyset $$ for each $i\in \{1,\dots ,n\}$ and $$[-2,2]^{n}
\subseteq f^{-1} (U_{1} )\cup \dots \cup f^{-1} (U_{n} )\cup
f^{-1} (V_{1} )\cup \dots \cup f^{-1} (V_{n} ).$$ This is
impossible by Theorem \ref{glavni}. So (\ref{thm-C1}) holds.

That (\ref{thm-C2}) holds, it follows readily from property (5).
Note that (\ref{thm-C2}) implies $\Gamma \neq\emptyset $. Hence
$\tilde{e}(k)\leq \tilde{h}(k)$ for each $k\in \mathbb{N}$.

Let $g:\mathbb{N}\rightarrow \mathbb{N}$ be some computable
function such that $$\Lambda
_{g(k)}\approx_{2^{-k}}\bigcup_{\tilde{e}(k)\leq v_{1} ,\dots
,v_{n} \leq \tilde{h}(k)}J_{(\tilde{l}(k))_{v_{1} ,\dots ,v_{n}
}}.$$ Such a function can be obtained by defining
$\Phi:\mathbb{N}\rightarrow \mathcal{P}(\mathbb{N})$, $\Phi
(k)=\{\tau _{1} (((l)_{v_{1} ,\dots ,v_{n} })_{0})\mid
\tilde{e}(k)\leq v_{1} ,\dots ,v_{n} \leq \tilde{h}(k)\}$  and
taking a computable $g:\mathbb{N}\rightarrow \mathbb{N}$ such that
$\Phi (k)=[g(k)]$ for each $k\in \mathbb{N}$.

By (\ref{thm-C1}) and (\ref{thm-C2})
\begin{equation}\label{thm-11}
S'\cap J_{x'}\prec_{2^{-k}}\Lambda _{g(k)}\mbox{ and }\Lambda
_{g(k)}\prec_{2\cdot 2^{-k}}S
\end{equation}
 for each $k\in \mathbb{N}$.  Since $f(\langle -1,1\rangle ^{n} )$ is an open
set in $U$ and $U$ is open in $S$, $f(\langle -1,1\rangle ^{n} )$
is open in $S$ and by (\ref{thm-10}) we have $f(\langle
-1,1\rangle ^{n} )\subseteq S'$. Hence $S'$ is a neighborhood of
$x$ in $S$ and therefore  $S'\cap J_{x'}$ is a neighborhood of $x$
in $S$. By (\ref{thm-11}) $S'\cap J_{x'}$ is computable up to
$S$.\qed

For $n\in \mathbb{N}$, $n\geq 1$, let  $$\mathbb{H}^{n} =\{(x_{1}
,\dots ,x_{n} )\in \mathbb{R}^{n} \mid x_{n} \geq 0\}.$$ The
subset of $\mathbb{H}^{n} $ consisting of those points whose last
coordinate is zero is denoted by $\Bd \mathbb{H}^{n} $.

\begin{thm} \label{lok-izr-eukld-skupa-mnsrubom}
Let $(X,d,\alpha )$ be a computable metric space and let $S$  and
$T$ be semi-computable compact sets in this space such that $T
\subseteq S$. Suppose $x\in S$ is a point which has a neighborhood
$U$ (in $S$) with the following property: there exists a
homeomorphism $f:\mathbb{H}^{n}\rightarrow U$  such that $f(\Bd
\mathbb{H}^{n})=U\cap T$.  Then there exists a neighborhood of $x$
in $S$ which is computable up to $S$.
\end{thm}
\proof If $f^{-1} (x)\in \mathbb{H}^{n} \setminus (\Bd
\mathbb{H}^{n})$, then $x$ has a neighborhood homeomorphic to some
open ball in $\mathbb{R}^{n}$, hence homeomorphic to
$\mathbb{R}^{n}$ and the claim of the theorem follows from Theorem
\ref{lok-izr-eukld-skupa}.

Suppose now that $f^{-1} (x)\in \Bd \mathbb{H}^{n}$. As in the
proof of Theorem \ref{lok-izr-eukld-skupa}, we may assume that $U$
is open is $S$ and $f(0)=x$. As before  we conclude that
 there exists $m_{0}
\in \mathbb{N}$ such that
$$S\setminus f(\langle -4,4\rangle ^{n} \cap \mathbb{H}^{n})\subseteq J_{m_{0}
}\mbox{ and }J_{m_{0} }\cap f([-2,2]^{n}\cap \mathbb{H}^{n}
)=\emptyset .$$ Let $S'=S\setminus J_{m_{0} }$. By Lemma
\ref{S-bez-Jj0} $S'$ is a semi-computable compact set in
$(X,d,\alpha )$ and
\begin{equation}\label{thm-10-mnsr}
f([-2,2]^{n}\cap \mathbb{H}^{n} )\subseteq S'\subseteq
f([-4,4]^{n}\cap \mathbb{H}^{n} ).
\end{equation}
Let $T'=T\setminus J_{m_{0} }$. Then $T'$ is semi-computable and
\begin{equation}\label{thm-11-mnsr}
f([-2,2]^{n}\cap (\Bd \mathbb{H}^{n}) )\subseteq T'\subseteq
f([-4,4]^{n}\cap (\Bd \mathbb{H}^{n}) ).
\end{equation}

 For $i\in
\{1,\dots ,n¸-1\}$ let $$C_{i} =\{(x_{1} ,\dots ,x_{n} )\in
[-4,4]^{n}\cap \mathbb{H}^{n} \mid x_{i} \geq -1\},~~D_{i}
=\{(x_{1} ,\dots ,x_{n} )\in [-4,4]^{n} \cap \mathbb{H}^{n}\mid
x_{i} \leq 1\}.$$ Let $$D_{n} =\{(x_{1} ,\dots ,x_{n} )\in
[-4,4]^{n} \cap \mathbb{H}^{n}\mid x_{n} \leq 1\}.$$ For $i\in
\{1,\dots ,n-1\}$ let
$$\tilde{A} _{i} =\{(x_{1} ,\dots ,x_{n}
)\in [-2,2]^{n}\cap \mathbb{H}^{n}  \mid x_{i} =-2\},$$
$$ \tilde{B} _{i} =\{(x_{1} ,\dots
,x_{n} )\in [-2,2]^{n}\cap \mathbb{H}^{n} \mid x_{i} =2\}.$$ Let
$$\tilde{A} _{n} =\{(x_{1} ,\dots ,x_{n}
)\in [-2,2]^{n}\cap \mathbb{H}^{n}  \mid x_{n} =0\},~~\tilde{B}
_{n} =\{(x_{1} ,\dots ,x_{n} )\in [-2,2]^{n}\cap \mathbb{H}^{n}
\mid x_{n} =2\}.$$

For $m\in \mathbb{N}$ let $E^{m} :\mathbb{N}_{8m+7} ^{n}
\rightarrow \mathcal{P}([-4,4]^{n}\cap \mathbb{H}^{n}   )$ be
defined so that $E^{m}_{i_{1} ,\dots ,i_{n} }$ is equal to
$$\left[-4+\frac{i_{1}
}{m+1},-4+\frac{i_{1} +1}{m+1}\right]\times \dots \times
\left[-4+\frac{i_{n-1} }{m+1},-4+\frac{i_{n-1}
+1}{m+1}\right]\times \left[\frac{i_{n} }{2m+2},\frac{i_{n}
+1}{2m+2}\right].$$ Then $E^{m} $ is a compact $n$-chain in
$[-4,4]^{n}\cap \mathbb{H}^{n} $ which covers $[-4,4]^{n}\cap
\mathbb{H}^{n} $. Let $F^{m} :\mathbb{N}_{8m+7} ^{n} \rightarrow
\mathcal{P}(X  )$ be defined by $F^{m} (v)=f(E^{m} (v))$,
$v\in\mathbb{N}_{8m+7} ^{n} $. Then $F^{m} $ is a compact
$n$-chain in $(X,d)$ which covers $f([-4,4]^{n}\cap
\mathbb{H}^{n}) $.  Note that the lower boundary of $F^{m} $
covers $f([-4,4]^{n} \cap (\Bd \mathbb{H}^{n}))$. Therefore by
(\ref{thm-11-mnsr}) the lower boundary of $F^{m} $ covers $T'$.

Exactly as in the proof of Theorem \ref{lok-izr-eukld-skupa} we
get numbers $a_{1} ,\dots ,a_{n} ,b_{1} ,\dots ,b_{n} ,x'\in
\mathbb{N}$ so that, for each $i\in \{1,\dots ,n\}$,
$f(\tilde{A}_{i} )\subseteq J_{a_{i} }$, $f(\tilde{B}_{i}
)\subseteq J_{b_{i} }$, $x\in J_{x'}$ and so that for each $k\in
\mathbb{N}$ there exist $l,e,h,u\in \mathbb{N}$ with the following
properties:
\begin{enumerate}%[(1)]
\item $\mathcal{H}_{l} $ covers $S'$;

\item the lower boundary of $\mathcal{H}_{l} $ covers $T'$;

\item $\mathcal{H}_{l} $ is a formal $n$-chain;

\item $J_{(l)_{v_{1} ,\dots ,v_{n} }}$ and $J_{a_{i} }$ are
formally disjoint if $1\leq i\leq n-1$, $0\leq v_{1} ,\dots ,v_{n}
\leq \widehat{l}$ and  $v_{i} \geq e$;

\item $J_{(l)_{v_{1} ,\dots ,v_{n} }}$ and $J_{b_{i} }$ are
formally disjoint if $1\leq i\leq n-1$, $0\leq v_{1} ,\dots ,v_{n}
\leq \widehat{l}$ and  $v_{i} \leq  h$;

\item $J_{(l)_{v_{1} ,\dots ,v_{n} }}$ and $J_{b_{n} }$ are
formally disjoint if $0\leq v_{1} ,\dots ,v_{n} \leq \widehat{l}$
and  $v_{n} \leq  u$;

\item $J_{(l)_{v_{1} ,\dots ,v_{n} }}$ and $J_{x' }$ are formally
disjoint if $1\leq i\leq n-1$, $0\leq v_{1} ,\dots ,v_{n} \leq
\widehat{l}$ and  ($v_{i}  <e$ or $v_{i} >h$);

\item $J_{(l)_{v_{1} ,\dots ,v_{n} }}$ and $J_{x' }$ are formally
disjoint if $0\leq v_{1} ,\dots ,v_{n} \leq \widehat{l}$ and
$v_{n} >u$;

\item $\fmesh (l)<2^{-k}$.

\end{enumerate}

The set $\Omega $ of all $(k,l,e,h,u)\in \mathbb{N}^{5}$ such that
(1)-(9) hold is c.e. Therefore there exist computable functions
 $\tilde{l},\tilde{e},\tilde{h},
\tilde{u}:\mathbb{N}\rightarrow \mathbb{N}$  such that
$(k,\tilde{l}(k),\tilde{e}(k),\tilde{h}(k),\tilde{u}(k))\in \Omega
$ for each $k\in \mathbb{N}$. Let $k\in \mathbb{N}$ and let
$l=\tilde{l}(k)$, $e=\tilde{e}(k)$, $h=\tilde{h}(k)$ and
$u=\tilde{u}(k)$. Let $\Gamma$ be the set of all $(v_{1} ,\dots
,v_{n} )\in \mathbb{N}^{n} _{\widehat{l}}$ such that $v_{n} \leq
u$ and $e\leq v_{i}\leq h$ for each $i\in \{1,\dots ,n-1\}$. We
claim that
\begin{equation}\label{thm-C1-mnsr}
J_{(l)_{v}}\cap f([-2,2]^{n}\cap \mathbb{H}^{n}  )\neq \emptyset
\mbox{ for all }v\in \Gamma,
\end{equation}
and
\begin{equation}\label{thm-C2-mnsr}
S'\cap J_{x' }\subseteq \bigcup_{v\in \Gamma }J_{(l)_{v}}.
\end{equation}
Suppose $J_{(l)_{v}}\cap f([-2,2]^{n} )=\emptyset$ for some
$v=(v_{1} ,\dots ,v_{n} )\in \Gamma$ such that $v_{n} =0$. For
each $i\in \{1,\dots ,n-1\}$ let $U_{i} $ be the union of all
$J_{(l)_{w_{1} ,\dots ,w_{n} }}$ such that $(w_{1} ,\dots ,w_{n}
)\in \mathbb{N}^{n} _{\widehat{l}}$, $w_{n} =0$ and  $w_{i} <v_{i}
$ and let $V_{i} $ be the union of all $J_{(l)_{w_{1} ,\dots
,w_{n} }}$ such that $(w_{1} ,\dots ,w_{n} )\in \mathbb{N}^{n}
_{\widehat{l}}$, $w_{n} =0$  and $w_{i}
>v_{i} $. It follows from (2) and (\ref{thm-11-mnsr}) that $$f([-2,2]^{n}\cap (\Bd \mathbb{H}^{n}) )\subseteq (U_{1} \cup \dots \cup U_{n-1}
)\cup (V_{1} \cup \dots \cup V_{n-1} ).$$ Since $\mathcal{H}_{l}$
is a chain, $U_{i} \cap V_{i} =\emptyset $ for each $i\in
\{1,\dots ,n-1\}$. Since $e\leq v_{i} \leq h$, by properties (4)
and (5) we have $U_{i} \cap J_{b_{i} }=\emptyset $ and $V_{i} \cap
J_{a_{i} }=\emptyset $. Hence $U_{i} \cap f(\tilde{B}_{i}
)=\emptyset $ and $V_{i} \cap f(\tilde{A}_{i} )=\emptyset $.
Therefore
\begin{equation}\label{thm-12-mnsr}
f^{-1} (U_{i} )\cap \tilde{B}_{i} =\emptyset,\mbox{ }f^{-1} (V_{i}
)\cap \tilde{A}_{i} =\emptyset\mbox{ and }f^{-1} (U_{i} )\cap
f^{-1} (V_{i} )=\emptyset
\end{equation}
 for each $i\in \{1,\dots ,n-1\}$ and
$$[-2,2]^{n}\cap (\Bd \mathbb{H}^{n})\subseteq f^{-1} (U_{1} )\cup \dots \cup f^{-1}
(U_{n-1} )\cup f^{-1} (V_{1} )\cup \dots \cup f^{-1} (V_{n-1} ).$$
This is impossible by Theorem \ref{glavni} (formally, we take the
function $\gamma :\mathbb{R}^{n-1}\rightarrow \mathbb{H}^{n}$,
$\gamma (z)= (z,0)$ and then for each $i\in \{1,\dots ,n-1\}$ we
get from (\ref{thm-12-mnsr}) that
$$\gamma ^{-1} (f^{-1} (U_{i} ))\cap B^{n-1}_{i} =\emptyset,~~\gamma ^{-1} (f^{-1} (V_{i}
))\cap A^{n-1}_{i} =\emptyset,~~\gamma ^{-1} (f^{-1} (U_{i} ))\cap
\gamma ^{-1} (f^{-1} (V_{i} ))=\emptyset$$ which is impossible
since the sets $\gamma ^{-1} (f^{-1} (U_{1} ))$, \dots, $\gamma
^{-1} (f^{-1} (U_{n-1} ))$, $\gamma ^{-1} (f^{-1} (V_{1} ))$,
\dots, $\gamma ^{-1} (f^{-1} (V_{n-1} ))$ are open in
$\mathbb{R}^{n-1}$ and cover $[-2,2]^{n-1}=I^{n-1}$.)

So $J_{(l)_{v}}\cap f([-2,2]^{n} )\neq\emptyset$ for each
$v=(v_{1} ,\dots ,v_{n} )\in \Gamma$ such that $v_{n} =0$.

Let $v=(v_{1} ,\dots ,v_{n} )\in \Gamma$ be such that $v_{n} >0$.
Suppose $J_{(l)_{v}}\cap f([-2,2]^{n} )=\emptyset$.

For $i\in \{1,\dots ,n\}$ we define $U_{i} $ as the union of all
$J_{(l)_{w_{1} ,\dots ,w_{n} }}$ such that $(w_{1} ,\dots ,w_{n}
)\in \mathbb{N}^{n} _{\widehat{l}}$ and $w_{i} <v_{i} $ and we
define $V_{i} $ as the union of all $J_{(l)_{w_{1} ,\dots ,w_{n}
}}$ such that $(w_{1} ,\dots ,w_{n} )\in \mathbb{N}^{n}
_{\widehat{l}}$ and $w_{i}
>v_{i} $. As before we conclude that
$$f([-2,2]^{n}\cap \mathbb{H}^{n}  )\subseteq (U_{1} \cup \dots \cup U_{n-1}
)\cup (V_{1} \cup \dots \cup V_{n-1} )$$ and $$U_{i} \cap V_{i}
=\emptyset,~~ U_{i} \cap f(\tilde{B}_{i} )=\emptyset,~~ V_{i} \cap
f(\tilde{A}_{i} )=\emptyset$$ for each $i\in \{1,\dots,n-1\}$. We
also have $U_{n} \cap V_{n} =\emptyset$. Since $v_{n} \leq u$,
property (6) implies $U_{n} \cap f(\tilde{B}_{n} )=\emptyset$. On
the other hand, by (\ref{thm-11-mnsr}) we have $f(\tilde{A}_{n}
)\subseteq T'$ and this, together with property (2), implies
$f(\tilde{A}_{n} )\subseteq U_{n} $. Therefore  $f(\tilde{A}_{n}
)\cap V_{n}=\emptyset$.

We have the following conclusion: for each $i\in \{i,\dots ,n\}$
$$f^{-1} (U_{i} )\cap \tilde{B}_{i} =\emptyset,\mbox{ }f^{-1} (V_{i} )\cap
\tilde{A}_{i} =\emptyset,\mbox{ }f^{-1} (U_{i} )\cap f^{-1} (V_{i}
)=\emptyset $$ and $$[-2,2]^{n}\cap \mathbb{H}^{n} \subseteq
f^{-1} (U_{1} )\cup \dots \cup f^{-1} (U_{n} )\cup f^{-1} (V_{1}
)\cup \dots \cup f^{-1} (V_{n} ).$$ To get a contradiction with
Theorem \ref{glavni} we define the function $\gamma
:\mathbb{R}^{n} \rightarrow \mathbb{H}^{n}$ by $$\gamma (z_{1}
,\dots ,z_{n} )=\left\{\begin{tabular}{l}
             $(z_{1} ,\dots ,z_{n-1},\frac{z_{n} +2}{2})$, if $z_{n} \geq -2$  \\
             $ (z_{1} ,\dots ,z_{n-1},0)$, if $z_{n} \leq -2.$
              \end{tabular} \right.$$
This is clearly a continuous function and we have $\gamma
([-2,2]^{n} )=[-2,2]^{n}\cap \mathbb{H}^{n}$. It follows
$$[-2,2]^{n}\subseteq
\gamma ^{-1} (f^{-1} (U_{1} ))\cup \dots \cup \gamma ^{-1} (f^{-1}
(U_{n} ))\cup \gamma ^{-1} (f^{-1} (V_{1} ))\cup \dots \cup \gamma
^{-1} (f^{-1} (V_{n} )).$$ For each $i\in \{1,\dots ,n\}$ clearly
$\gamma ^{-1} (f^{-1} (U_{i} ))\cap \gamma ^{-1} (f^{-1} (V_{i}
))=\emptyset$,
$$\gamma ^{-1} (f^{-1} (U_{i} ))\cap \gamma ^{-1} (\tilde{B}_{i}) =\emptyset\mbox{ and }\gamma ^{-1} (f^{-1} (V_{i} ))\cap
\gamma ^{-1} (\tilde{A}_{i}) =\emptyset .$$ It is immediate from
the definition of $\gamma $ that $A_{i} ^{n} \subseteq \gamma
^{-1} (\tilde{A}_{i} )$ and $B_{i} ^{n} \subseteq \gamma ^{-1}
(\tilde{B}_{i} )$ for each $i\in \{1,\dots ,n\}$, therefore
$\gamma ^{-1} (f^{-1} (U_{i} ))\cap B^{n} _{i} =\emptyset$ and
$\gamma ^{-1} (f^{-1} (V_{i} ))\cap A^{n} _{i} =\emptyset $. This
altogether contradicts Theorem \ref{glavni}.
 So we finally conclude that (\ref{thm-C1-mnsr}) holds.

That (\ref{thm-C2-mnsr}) holds, it follows  from properties (7)
and (8). We note as before that $\Gamma \neq\emptyset $ and
therefore $\tilde{e}(k)\leq \tilde{h}(k)$ for each $k\in
\mathbb{N}$.

Let $g:\mathbb{N}\rightarrow \mathbb{N}$ be some computable
function such that $$\Lambda
_{g(k)}\approx_{2^{-k}}\bigcup_{\tilde{e}(k)\leq v_{1} ,\dots
,v_{n-1} \leq \tilde{h}(k),~0\leq v_{n} \leq
\tilde{u}(k)}J_{(\tilde{l}(k))_{v_{1} ,\dots ,v_{n} }}.$$

By (\ref{thm-C1-mnsr}) and (\ref{thm-C2-mnsr})
$$S'\cap
J_{x'}\prec_{2^{-k}}\Lambda _{g(k)}\mbox{ and }\Lambda
_{g(k)}\prec_{2\cdot 2^{-k}}S$$
 for each $k\in \mathbb{N}$. Since $S'\cap J_{x'}$ is a neighborhood of $x$ in $S$,
 the claim of the theorem follows. \qed

\section{Semi-computable compact manifolds with boundary} \label{sect-6}

Let $n\in \mathbb{N}$, $n\geq 1$. A nonempty topological space $X$
is said to be an $n$-\textbf{manifold} if each point of $X$ has a
neighborhood which is homeomorphic to $\mathbb{R}^{n} $.  The
following are some examples of $n$-manifolds: Euclidean space
$\mathbb{R}^{n} $, the unit sphere  $$S^{n} =\{x\in
\mathbb{R}^{n+1}\mid \|x\|=1\}$$ in $\mathbb{R}^{n+1}$, the real
projective space $P^{n} =\{\{x,-x\}\mid x\in S^{n} \}$ (with
topology on $P^{n} $ defined by $U\subseteq P^{n} $ is open if the
union of the family $U$ is open in $S^{n}$, see \cite{cv}). If $X$
is an $n$-manifold and $Y$ is an $m$-manifold, then $X\times Y$ is
an $n\cdot m$-manifold. Therefore, the space $S^{1}\times \dots
\times S^{1}$ ($n$-times) is an $n$-dimensional manifold. The
\textbf{torus} is the space $S^{1}\times S^{1}$, hence it is a
$2$-manifold.

A nonempty topological space $X$ is said to be an
$n$-\textbf{manifold with boundary} if each point of $X$ has a
neighborhood which is homeomorphic to an open subset of
$\mathbb{H}^{n} $. If $X$ is an $n-$manifold with boundary, then
the subset of $X$ consisting of all points $x\in X$ with the
property that there exists a neighborhood $N$ of $x$, an open
subset $U$ of $\mathbb{H}^{n}$ and a homeomorphism $f:N\rightarrow
U$ such that $f(x)\in \Bd \mathbb{H}^{n}$ is called the
\textbf{boundary} of $X$. We denote the boundary of $X$ by
$\partial X$. It can be shown (see \cite{mu}) that if $x\in
\partial X$, then every homeomorphism between a neighborhood of
$x$ in $X$ and an open subset of $\mathbb{H}^{n}$ maps $x$ to a
point in $\Bd \mathbb{H}^{n}$. It is interesting to remark here
that if a point in some topological space has neighborhoods $N_{1}
$ and $N_{2} $ which are homeomorphic to open subsets of
$\mathbb{H}^{n}$ and $\mathbb{H}^{m} $ respectively, then $n=m$;
this is a consequence of Invariance of domain theorem, see
\cite{mu}. In particular, the number $n$, called the dimension of
$X$, is uniquely determined by $X$.

If $U$ is an open set in $\mathbb{H}^{n} $ and $z\in U$ is a point
such that $z\in \mathbb{H}^{n}\setminus (\Bd \mathbb{H}^{n})$,
then there exists an open ball in $\mathbb{R}^{n}$ centered at $x$
which is contained in $U$. Furthermore, each open ball in
$\mathbb{R}^{n} $ is homeomorphic to $\mathbb{R}^{n} $. Therefore,
if $X$ is an $n-$manifold with boundary and $x\in X\setminus
\partial X$, then $x$ has an open neighborhood in $X$ homeomorphic to
$\mathbb{R}^{n} $. On the other hand, each open ball in
$\mathbb{H}^{n} $ centered at a point in $\Bd \mathbb{H}^{n}$ is
homeomorphic to $\mathbb{H}^{n} $. It follows from this that each
point in $\partial X$ has an open neighborhood in $X$ homeomorphic
to $\mathbb{H}^{n} $.

If $X$ is a manifold, then clearly $X$ is a manifold with boundary
and $\partial X=\emptyset $. In fact, if $X$ is a manifold with
boundary, then $X$ is a manifold if and only if $\partial
X=\emptyset $.

If $X$ is a manifold with boundary and $Y$ is homeomorphic to $X$,
then clearly $Y$ is also a manifold with boundary and $\partial
Y=f(\partial X)$, where $f:X\rightarrow Y$ is a homeomorphism.
Furthermore, if $X$ and $Y$ are manifolds with boundary, then
$X\times Y$ is a manifold with boundary and the point $(x,y)$ is
in its boundary if and only if $x\in \partial X$ or $y\in \partial
Y$.

It is not hard to check (see \cite{mu}) that the closed unit ball
$B^{n} $ in $\mathbb{R}^{n} $ is a manifold with boundary and
$\partial B^{n} =S^{n-1}$. For example $[0,1]$ is a manifold with
boundary and $\partial [0,1]=\{0,1\}$. Therefore, since $S^{1}$ is
a manifold,  $S^{1}\times [0,1]$ is a manifold with boundary and
$\partial (S^{1}\times [0,1])=S^{1}\times \{0,1\}$. (Note that the
topological boundary of $S^{1}\times [0,1]$ in $\mathbb{R}^{3}$
differs from $\partial (S^{1}\times [0,1])$.)

\begin{thm} \label{comp-comp-man}
Let $(X,d,\alpha )$ be a computable metric space and let
$S\subseteq X$. Suppose $S$, as a subspace of $(X,d)$, is a
manifold with boundary and suppose $S$ and $\partial S$ are
semi-computable compact sets in $(X,d,\alpha )$. Then $S$ is a
computable compact set in $(X,d,\alpha )$.
\end{thm}
\proof Let $n\geq 1$ be such that $S$ is an $n$-manifold with
boundary. If $x\in S\setminus \partial S$, then $x$ has a
neighborhood homeomorphic to $\mathbb{R}^{n} $ and  by Theorem
\ref{lok-izr-eukld-skupa} there exists a neighborhood of $x$ in
$S$ which is computable up to $S$. On the other hand, if $x\in
\partial S$, then there exists an open neighborhood $U$ of $x$ in
$S$ and a homeomorphism $f:\mathbb{H}^{n} \rightarrow U$. It is
clear that $f(\Bd \mathbb{H}^{n})\subseteq U\cap
\partial S$ and since for each $z\in \partial S$ every homeomorphism between a neighborhood
of $z$ in $S$ and an open subset of $\mathbb{H}^{n}$ maps $z$ to a
point in $\Bd \mathbb{H}^{n}$, we have $f(\Bd \mathbb{H}^{n})=
U\cap \partial S$. It follows from Theorem
\ref{lok-izr-eukld-skupa-mnsrubom} that $x$ has a neighborhood  in
$S$ which is computable up to $S$.

Hence each $x\in S$ has a neighborhood $N_{x}$ in $S$ which is
computable up to $S$. Let $\mathcal{U}=\{\Int_{S}N_{x}\mid x\in
S\}$. Then $\mathcal{U}$ is an open cover of $S$. Since $S$ is
compact, there exists a finite subcover of $\mathcal{U}$ for $S$,
hence there exist $x_{1} ,\dots ,x_{k} \in S$ such that
$S=N_{x_{1} }\cup \dots \cup N_{x_{k} }$. By Proposition
\ref{lokal-global} $S$ is a computable compact set. \qed

\begin{cor} \label{comp-comp-man-kor1}
Let $(X,d,\alpha )$ be a computable metric space and let $S$ be a
semi-computable compact set in this space which is a manifold with
boundary. Then the following implication holds: $$\partial S\mbox{
computable compact set}\Rightarrow S\mbox{ computable compact
set}.\eqno{\qEd}$$ 
\end{cor}

\begin{thm} \label{comp-comp-man-2}
Let $(X,d,\alpha )$ be a computable metric space. Let $S$ be a
semi-computable compact set in this space such that $S$, as a
subspace of $(X,d)$, is a manifold. Then $S$ is computable compact
in $(X,d,\alpha )$. \qed
\end{thm}

In general, if $X$ is an $n-$manifold with boundary and $\partial
X\neq\emptyset $, then $\partial X$ is an $(n-1)-$manifold.
Therefore if $S$ is a manifold with boundary in $(X,d,\alpha )$,
then $\partial S$ is a manifold (boundaryless) and we have, by
Theorem \ref{comp-comp-man-2}, that $\partial S$ is
semi-computable compact if and only if $\partial S$ is computable
compact. This means that Corollary \ref{comp-comp-man-kor1} is an
equivalent formulation of Theorem \ref{comp-comp-man}.

\begin{thm} \label{comp-comp-man-kor}
Let $(X,d,\alpha )$ be a computable metric space which is locally
computable. Let $S$ be a co-computably enumerable closed set in
this space such that $S$, as a subspace of $(X,d)$, is a compact
manifold with boundary. Then the following implication holds:
$$\partial S\mbox{ computable closed set }\Rightarrow S\mbox{ computable closed set}.\eqno{\qEd}$$
\end{thm}

\subsection{Application: regular level sets} Suppose that
$f:\mathbb{R}^{n} \rightarrow \mathbb{R}^{m} $ is a function of
class C$^{1}$ such that $y\in \mathbb{R}^{m} $ is a regular value
of $f$, which means that the differential $D(f)(x):\mathbb{R}^{n}
\rightarrow \mathbb{R}^{m} $ of $f$ in $x$ is surjective for each
$x\in f^{-1} \{y\}$. Suppose $f^{-1} \{y\}\neq\emptyset$. Then it
is known from differential topology (see \cite{gil-pol}) that
$f^{-1} \{y\}$ is an $(n-m)$-manifold.

Suppose now additionally that $f$ is computable and  $f^{-1}
\{y\}$ is a bounded set. In general, if $g:\mathbb{R}^{n}
\rightarrow \mathbb{R}$ is a computable function, then $g^{-1}
\{0\}$ is a co-c.e.\ closed set in $\mathbb{R}^{n} $. Therefore
$f^{-1} \{y\}$ is a co-c.e.\ closed set and, by Theorem
\ref{comp-comp-man-kor}, we conclude the following.

\begin{cor} \label{level-sets}
If $f:\mathbb{R}^{n} \rightarrow \mathbb{R}^{m} $ is a computable
function of class C$^{1}$ and $y\in \mathbb{R}^{m} $ a regular
value of $f$ such that $f^{-1}\{y\}$ is bounded, then the set
$f^{-1}\{y\}$ is computable. \qed
\end{cor}

\begin{exa}
(i) Let $S$ be the set of all $(x,y,z)\in \mathbb{R}^{3}$ such
that
$$x^{2}(1+e^{x})+y^{2}(1+e^{y})+z^{2}(1+e^{z})=1.$$
 We have $S=f^{-1} \{1\}$, where
$f:\mathbb{R}^{3}\rightarrow \mathbb{R}$ is defined in the obvious
way.  Since $\nabla f(x)=(\partial _{1}f(x),\partial _{2}f(x),
\partial _{3}f(x))=0$ only for the point $x=0$ which is not in $ f^{-1} \{1\}$, we have that
$1$ is a regular value for $f$. Hence $S$ is computable.
%In the
%same way we conclude that the set of all $(x,y,z)\in
%\mathbb{R}^{3}$ such that $$x^{2}+y^{2}(\sin x+2)+z^{4}e^{y+x}=3$$
%is computable.

(ii) Let $S$ be the set of all $(x_{1} ,x_{2} ,x_{3} ,x_{4} )\in
\mathbb{R}^{4}$ such that
$$x_{1} ^{2}+x_{2} ^{2}+x_{3} ^{2}+x_{4} ^{2}=1$$ and
$$\sin x_{1}+\sin x_{2}+\sin x_{3}+\sin x_{4}=1.$$
The intersection of two computable sets need not be computable in
general and therefore we cannot get that $S$ is computable only by
proving that the sets defined by above equations are computable.
We have $S=f^{-1} \{(1,1)\}$, where $f:\mathbb{R}^{3}\rightarrow
\mathbb{R}^{2}$ is defined in the obvious way.  Let $f_{1} ,f_{2}
$ be the component functions of $f$. It is easy to check that the
vectors $\nabla f_{1} (x)$ and $\nabla f_{2} (x)$ are linearly
independent for each $x\in S$. By Corollary \ref{level-sets} $S$
is computable.
\end{exa}

\section*{Acknowledgements}
The author would like to thank the anonymous referees for their
careful work.

\end{document}